\documentclass[numberedappendix]{emulateapj}

\def\rmit#1{{\it #1}}              %% italics (RR mode, Kluwer)

\def\ie{\rmit{i.e.}}
\def\eg{\rmit{e.g.}}

\shorttitle{Waves below sunspots}

\shortauthors{Khomenko et al.}

\begin{document}

\title{Theoretical modeling of propagation of magneto-acoustic
       waves in magnetic regions below sunspots.}

\author{E. Khomenko\altaffilmark{1,2}, A. Kosovichev\altaffilmark{3}, M. Collados\altaffilmark{1},
         K. Parchevsky\altaffilmark{3} and V. Olshevsky\altaffilmark{2} }
\email{khomenko@iac.es}

\altaffiltext{1}{Instituto de Astrof\'{\i}sica de Canarias, 38205,
C/ V\'{\i}a L{\'a}ctea, s/n, Tenerife, Spain}
\altaffiltext{2}{Main Astronomical Observatory, NAS, 03680, Kyiv,
Ukraine} \altaffiltext{3}{Stanford University, 455 Via Palou,
Stanford, CA 94305, United States}

\begin{abstract}
We use 2D numerical simulations and eikonal approximation, to
study properties of MHD waves traveling below the solar surface
through the magnetic structure of sunspots.
We consider a series of magnetostatic models of sunspots of
different magnetic field strengths, from 10 Mm below the
photosphere to the low chromosphere. The purpose of these studies
is to quantify the effect of the magnetic field on local
helioseismology measurements by modeling waves excited by
sub-photospheric sources. Time-distance propagation diagrams and
wave travel times are calculated for models of various field
strength and compared to the non-magnetic case. The results
clearly indicate that the observed time-distance helioseismology
signals in sunspot regions correspond to fast MHD waves. The slow
MHD waves form a distinctly different pattern in the time-distance
diagram, which has not been detected in observations. The
numerical results are in good agreement with the solution in the
short-wavelength (eikonal) approximation, providing its
validation. The frequency dependence of the travel times is in a
good qualitative agreement with observations.
\end{abstract}


%%%%%%%%%%%%%%%%%%%%%%%%%%%%%%%%%%%%%%%%%%%%%%%%%%%%%%%%%%%%%%%%%%%%%%%%%%%

\section{Introduction}

Local helioseismology (time-distance helioseismology, acoustic
holography and ring-diagram analysis) provides valuable
information about the physical properties of solar sub-surface
layers \citep[\eg ][]{Duvall+etal1993, Kosovichev1999,
Kosovichev2002, Kosovichev+etal2000, Zhao+Kosovichev2003,
Braun+Lindsey2000, Hill1988, Habler+etal2000}. Time-distance
helioseismology \citep{Duvall+etal1993, Kosovichev+Duvall1997}
makes use of wave travel times measured for wave packets traveling
between various points on the solar surface through the interior.
By inversion of these measurements, variations of the wave speed
and velocities of mass flows can be recovered below the visible
solar surface.
Inversion results for the travel times have been obtained for
quiet Sun regions as well as for magnetic active regions including
sunspots. In most of these cases, the variations of the travel
times were assumed to be caused by mass flows and magneto-sonic
speed fluctuations along the wave paths.
It is known that sunspots possess strong magnetic fields with a
complicated structure in the visible layers of the Sun where the
Doppler measurements used by local helioseismology are taken
\citep{Solanki2003}. Consequently, such magnetic field can cause
important effects on helioseismic waves, beyond the perturbation
theories employed for helioseismic data analysis
\citep{Kosovichev+Duvall1997, Birch+Kosovichev2000,
Gizon+Birch2002}.
Magnetic field can change the acoustic excitation rate and
spectral properties of solar oscillations \citep[see the recent
work by][]{Jacoutot+etal2008};  produce new wave modes; change
wave propagation speeds; change the wave propagation paths and
reflection characteristics at the near-surface layers.

The influence of magnetic field on the interpretation of local
helioseismology measurements has not been fully explored.
Theoretical efforts have been made by \citet{Crouch+Cally2003},
\citet{Cally2005, Cally2006}, \citet{Schunker+Cally2006},
\citet{Cally+Goossens2008}, \citet{Schunker+etal2008}  to include
mode conversion and model waves in magnetized structures by means
of analytical and ray-path theories.
These studies confirm the potential importance of magnetic
effects. A more complete understanding of the problem can be
reached by direct forward modeling of helioseismic data, by
solving numerically the MHD equations in realistic magnetic
configurations \citep[\eg, ][]{Gizon+etal2006,
Khomenko+Collados2006, Hanasoge2008, Cameron+etal2007,
Parchevsky+Kosovichev2008}.
In these papers, the MHD equations, governing the problem, are
solved for magnetic field configurations resembling a sunspot or a
part of a sunspot.
The results have demonstrated how magnetic fields change the
speed, amplitude and spectral properties of the waves, and also
produce wave transformation and refraction in the sunspot
atmosphere.

Observational helioseismology over active regions has demonstrated
repeatedly  that waves travel faster through sunspots in
comparison with quiet Sun measurements \citep[\eg
][]{Duvall+etal1993, Braun1997}. This fact is still the subject of
present investigations and its explanation is still far from
clear. Although, given that sunspots possess an intense field at
photospheric levels, direct or indirect the influence of the
magnetic field is generally accepted \citep[\eg
][]{Hindman+etal1997}. At sub-photospheric levels the influence of
the magnetic field becomes less and less important with depth as
pressure forces dominate over magnetic forces. This is why the
phenomena related to magnetic field are referred to as ``surface
effects''. Explanations based on wave perturbations due to
scattering phase shifts \citep{Braun1997, Lindsey+Braun2005a,
Lindsey+Braun2005b} taking place within a few Mm below the
photosphere or local wave speed variations below sunspots
\citep{Kosovichev+etal2000} have been suggested to explain the
different travel times measured in sunspots and in the quiet
photosphere.

In this paper, we study wave properties and variations of travel
times  with the help of numerical simulations and Eikonal
solutions for wave propagation in sunspot-like magnetic structures
in stratified surface and sub-surface layers.
We  perform 2D numerical simulations of the waves produced by a
single source for a series of model sunspots
\citep{Khomenko+Collados2008}. We study various types of MHD waves
and their propagation properties numerically. In addition, for the
fast magneto-acoustic mode we obtain a solution in the eikonal
approximation, which is often important for correct interpretation
of helioseismic results  \citep{Gough2007}. Our numerical results
clearly indicate that the fast mode is the primary source of
helioseismic signals. We study the frequency dependence of the
wave travel times and compare the results of the eikonal approach
with our numerical simulations and observational data. Our results
provide insight into the properties of helioseismic waves in
magnetic regions and are important for interpretation of
time-distance helioseismology measurements.

\section{Numerical procedure}

We use the numerical simulation code described previously by
\citet{Khomenko+Collados2006, Khomenko+etal2008}. The code solves
the basic non-linear equations of the ideal MHD, written in
conservative form:
\begin{equation}
 \frac{\partial \rho }{\partial \it{t}} +  \vec{\nabla} \cdot
(\rho \vec{V})=  0 \,, \label{eq:den}
\end{equation}
\begin{equation}
\frac{\partial (\rho \vec{V}) }{\partial \it{t}} + \vec{\nabla}
\cdot [ \rho \vec{V}\vec{V} + (P + \frac{\vec{B}^2}{8\pi}) {\bf I}
- \frac{\vec{B}\vec{B}}{4\pi}]=\rho\vec{g} \,, \label{eq:mom}
\end{equation}
\begin{equation}
\frac{\partial E}{\partial \it{t}} + \vec{\nabla}\cdot[(E + P +
\frac{\vec{B}^2}{8\pi})\vec{V} - \vec{B}(\frac{\vec{B} \cdot
\vec{V}}{4\pi})] = \rho\vec{V}\cdot\vec{g} + \rho Q  \,,
\label{eq:ene}
\end{equation}
\begin{equation}
 \label{eq:ind}
\frac{\partial \vec{B}}{\partial \it{t}} =
\vec{\nabla}\times(\vec{V} \times \vec{B})  \,,
 \end{equation}
where ${\bf I}$ is the identity tensor and $E$ is the total
energy:
\begin{equation}
E=\frac{1}{2}\rho V^2 + \frac{P}{\gamma-1} + \frac{B^2}{8\pi} \,.
 \end{equation}
All other symbols have their usual meaning. Gravity has been taken
constant, with a value of 2.74$\times 10^4$ cm/sec$^2$.
The term describing radiative energy losses, $Q$, is set to zero
in the present study.
A Perfectly Matching Layer (PML) boundary condition
\citep{Berenger1994} is applied to all boundaries of the
simulation domain with the aim of minimizing wave reflection
effects at the boundaries. This performs rather well in our
simulations. The thickness of the PML layer in our simulations is
20 grid points (1 Mm).

The code solves the non-linear MHD equations for perturbations,
that are obtained by subtracting the equations of initial
magnetostatic equilibrium from  Eqs.~\ref{eq:den}-\ref{eq:ind}.
However, for the purpose of the present study, we keep the
perturbation amplitudes as low as possible, making the 2nd order
non-linear terms relatively unimportant. In this paper, we use the
2D version of the code.

The wave source is modeled as a time-dependent term
$\vec{S}(x,z,t)$ in the right hand side of the momentum equation
(Eq.~\ref{eq:mom}) in the same way as in the paper by
\citet{Parchevsky+Kosovichev2008}:
\begin{equation}
f(x,z,\tau)=\left\{
\begin{array}{ll}
\displaystyle
A\left[1-\frac{r^2}{R_{src}^2}\right]^2\left(1-2\tau^2\right)e^{-\tau^2} & \mbox{if } r\leq R_{src}\\
\displaystyle 0 & \mbox{if } r>R_{src},
\end{array}
\right.
\end{equation}
where $R_{src}$ is the source radius, taken to be 250 km; $r$ is
the distance from the source center,
$r=\sqrt{(x-x_{src})^2+(z-z_{src})^2}$. The parameter $\tau$ is
given by the equation
\begin{equation}\label{Eq:MHD_3Dsource}
\tau=\frac{\omega (t-t_0)}{2} - \pi, \qquad t_0\leq t\leq
t_0+\frac{4\pi}{\omega},
\end{equation}
where $\omega$ is the central, dominating, frequency of the source
and $t_0$ is the source start moment. In our model the source
represents the vertical force in the momentum equation,
$\vec{S}=(0,f)^T$. The central frequency of the source is 3.3 mHz.
In all the simulations described here, the center of the source is
located at depth $-200$ km, below the photospheric level of the
reference model S of \citet{Christensen-Dalsgaard+etal1996}.
The depth location of the source in our simulations corresponds to
the recent studies of the excitation of waves in magnetic regions
by \citet{Jacoutot+etal2008}.

The source generates a broad spectrum of acoustic and surface
gravity waves in the 5-minute range, resembling the wave
excitation in the real Sun. In the case of a sunspot model, these
modes are modified by the presence of magnetic field.

%%%%%%%%%%%%%%%%%%%%%%%%%%%%%%%%%%%%%%%%%%%%%%%%%%%%%%%%%%%%%%%%%%%%%%%%%%%%%%%%%%%%%%
\begin{figure}
\centering
\includegraphics[width=8cm]{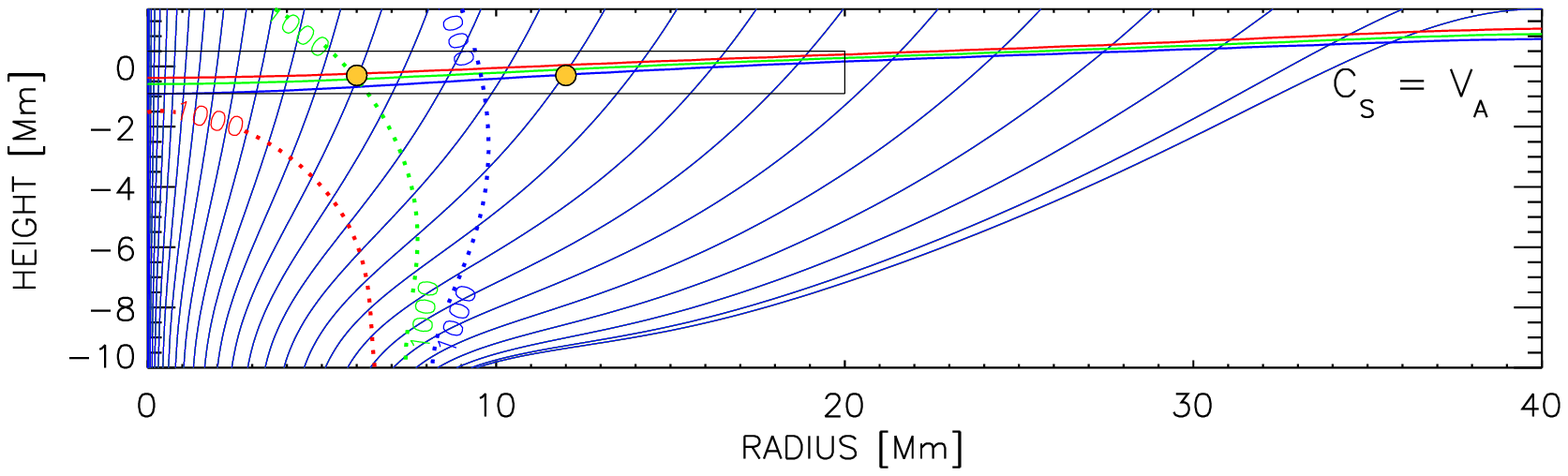}
\includegraphics[width=8cm]{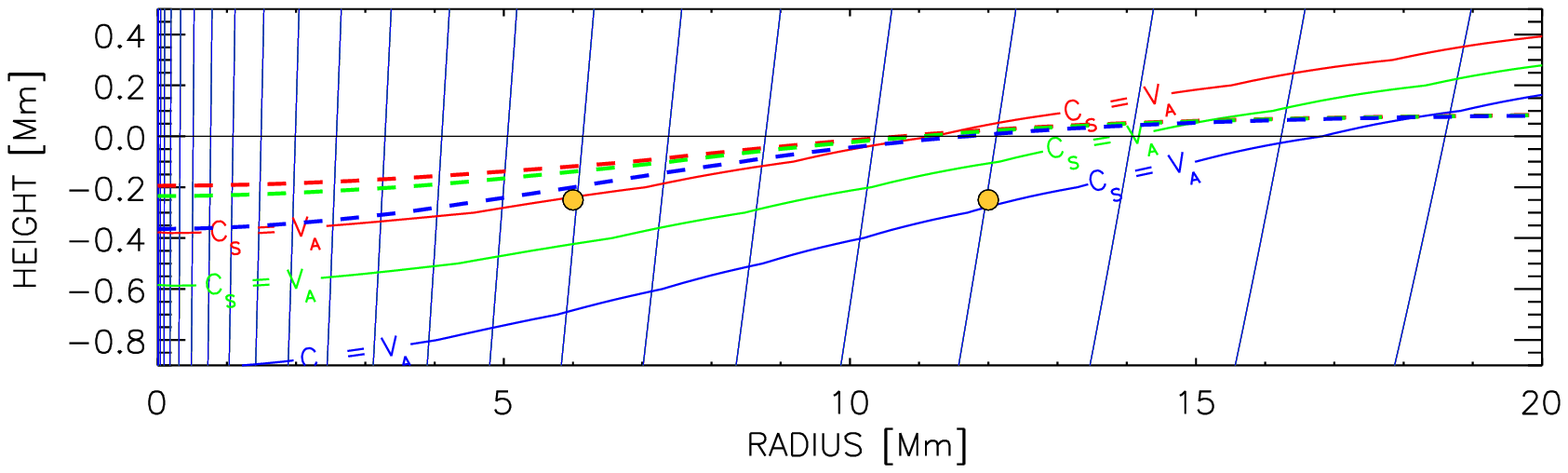}
\includegraphics[width=8cm]{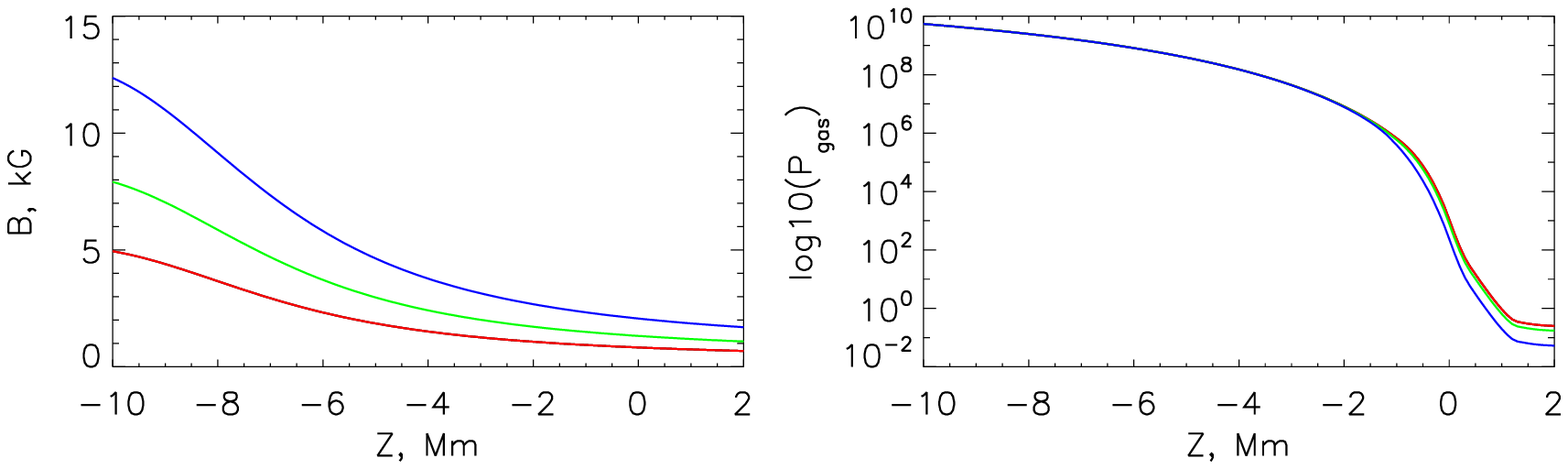}
\caption{Top panel: topology of the magnetic field lines for the
three sunspot models with $B_{\rm phot}$=0.9 kG (red lines), 1.5
kG (green lines) and 2.4 kG (blue lines). The horizontal solid
lines mark the levels where $c_S = v_A$. The contours of the
magnetic field strength of $B=1$ kG are shown by dotted curves for
each case. Two yellow dots mark the locations of the wave source
in the two sets of our simulations. Middle panel: enlarged view of
the upper part of the sunspot models marked by the box in the top
panel. The vertical axis has been expanded for better
visualization. Dashed lines are the contours of constant optical
depth log$\tau_5=-1.6$.  Bottom panels: the height dependence of
the magnetic field strength and the gas pressure at the axis for
these models. } \label{fig:spotmodel}
\end{figure}
%%%%%%%%%%%%%%%%%%%%%%%%%%%%%%%%%%%%%%%%%%%%%%%%%%%%%%%%%%%%%%%%%%%%%%%%%%%%%%%%%%%%%%%

%%%%%%%%%%%%%%%%%%%%%%%%%%%%%%%%%%%%%%%%%%%%%%%%%%%%%%%%%%%%%%%%%%%%%%%%%%%%%%%%%%%%%%
\begin{figure}
\centering
\includegraphics[width=8cm]{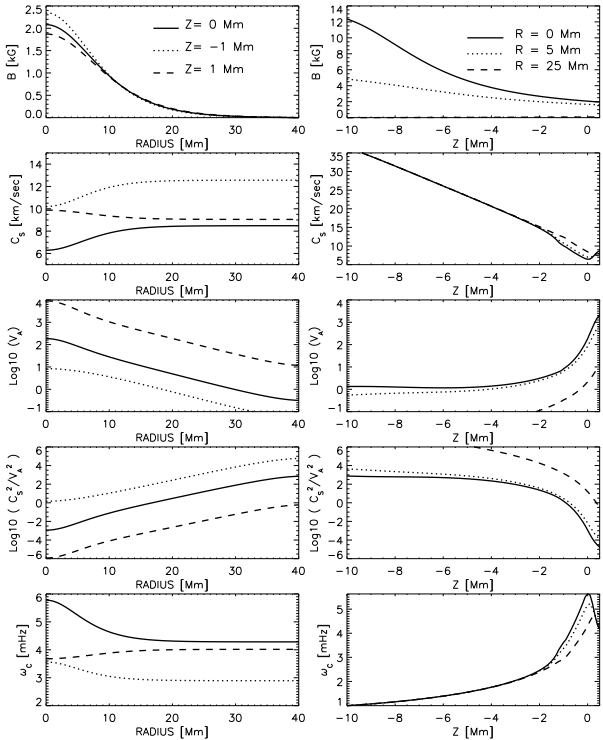}
\caption{The distribution with the radial distance (left panels)
and with the depth (right panels) of the magnetic field strength,
$B$; sound speed, $c_S$; Alfv\'en speed, $v_A$; their ratio,
$c_S^2/v_A^2$; and the acoustic cut-off frequency, $\omega_c$
(calculated in the assumption of an isothermal atmosphere) for the
$B_{\rm phot}$=2.4 kG sunspot. } \label{fig:spot4000hz}
\end{figure}
%%%%%%%%%%%%%%%%%%%%%%%%%%%%%%%%%%%%%%%%%%%%%%%%%%%%%%%%%%%%%%%%%%%%%%%%%%%%%%%%%%%%%%%

\section{Sunspot model and simulation setup}

The sunspot models are calculated using the method proposed in
\cite{Khomenko+Collados2008}. Here we use three magnetostatic
models of sunspots with similar magnetic field structure but
different values of the magnetic field strength. In this way we
are able to study the effects of the magnetic field strength on
the propagating waves apart from the effects of the magnetic field
inclination.
Some characteristics of these models are displayed in
Figures~\ref{fig:spotmodel} and \ref{fig:spot4000hz}. The
magneto-static equilibrium is calculated in the cylindrical
coordinates assuming azimuthal asymmetry and no twist of the
magnetic field lines. At large distances from the axis, the
sunspot models merge smoothly into the quiet Sun model, for which
we take the model S of \citet{Christensen-Dalsgaard+etal1996},
modified to make it stable with respect to convective motions
\citep[see][]{Parchevsky+Kosovichev2007b}.
The sound speed variations along the radial distance are evident
from Fig.~\ref{fig:spot4000hz}.  The sunspot models have a
low-temperature zone below the surface down to, about, $-2$ Mm
depth. Below this depth the temperature gradient in the horizontal
direction is small and no high-temperature zone is introduced in
the present study.
Note that the ratio between the acoustic and the Alfv\'en speeds
squared changes by orders of magnitude with height, and that the
gradient of this ratio is largest in the surface layers.

For the wave simulations, the models are transformed from
cylindrical coordinates $(r, \phi, z)$ to 3D cartesian
coordinates. In a general case, all three components of the
magnetic field vector are non-zero in the cartesian coordinates.
In the case of the 2D simulations, we chose a slice where the
$B_y$ component of the field is equal to zero and calculate the
wave propagation in this plane. By adopting this approach we
exclude Alfv\'en waves from consideration.

The simulation domain is 40 Mm in the horizontal direction and
10.5 Mm in the vertical direction. The sunspot axis is located at
the center of the simulation domain, and the upper 0.5 Mm of the
domain are located above $Z=0$ km height. Above $Z= 0.5$ Mm we
introduce an isothermal PML layer, 1 Mm thick.  The grid
resolution of the simulations is 50 km in the vertical and 100 km
in the horizontal directions. The temporal resolution depends on
the CFL numerical stability condition, limited by the rapid growth
of the Alfv\'en speed, $v_A$, with height, changing typically
between 0.01 and 0.08 seconds. For each of the three sunspot
models we perform two simulation runs with the wave source located
at $X_0=$ 6 Mm and 12 Mm to the left of the sunspot axis.

In addition to the simulations of waves in the sunspot models, we
have also performed a reference simulation for a model without
magnetic field and without horizontal variations in thermodynamic
parameters. The vertical stratification of the thermodynamic
parameters in this case corresponds to the modified model S.

Several recent  works on helioseismology point out the presence of
the so-called ``surface effects'' in the measurements in active
regions \citep[see \eg\ ][]{Lindsey+Braun2005a,
Lindsey+Braun2005b, Schunker+etal2005, Zhao+Kosovichev2006,
Rajaguru+etal2007}. Due to the Wilson depression in sunspots the
formation heights of spectral lines are altered introducing
additional phase shifts into the velocity measurements, as
compared to non-magnetic regions. Our sunspot models describe the
Wilson depression effect since the whole atmosphere inside the
magnetic concentration had been shifted downwards. In order to
take these ``surface effects'' into account,  we have computed the
optical depth scale $\tau_5$ in our sunspot models. This has been
done with the help of SIR code
\citep{RuizCobo+delToroIniesta1992}. As expected, the $\tau_5$
scale varies with radial distance from sunspot center.  This
variation can be appreciated at the middle panel of
Fig.~\ref{fig:spotmodel}, where the contours of the constant
optical depth log$\tau_5=-1.6$ are given. The optical depth
log$\tau_5=-1.6$ represents the layer where typical photospheric
spectral lines are formed. In the rest of the paper we call
``photospheric velocity'' the velocity measured at
log$\tau_5=-1.6$ level. Note that the height variations in the
position of this layer can be as large as 0.2--0.4 Mm, depending
on the distance from the axis in the sunspot models. In the
non-magnetic simulation the ``photospheric velocity'' at
log$\tau_5=-1.6$ always corresponds to the same height, as there
is no Wilson depression.

\section{Wave field from a single source}

In this section we describe the results of the simulations for the
three model sunspots and the quiet Sun model for two positions of
the oscillation source. In the first series of simulations, the
source is located in the ``umbral'' part of the model at a
distance of 6 Mm from the sunspot axis, where the field is
relatively strong. In this region the ratio $c_S^2/v_A^2$ is equal
to 1.4, 0.46 and 0.11 in the three models with $B_{\rm phot}=$
0.9, 1.5 and 2.4 kG, correspondingly. Thus, the wave source is
located in the region dominated by the magnetic field in the last
two models.
In the second series, the source is located in the ``penumbral''
part of the model, at a distance of 12 Mm from the axis, where the
field is weaker. There the parameter  $c_S^2/v_A^2$ is equal to
8.2, 3.1 and 1.1 in the three models. In this case the waves are
excited in the acoustically dominated region.
Placing the wave source at these two positions allows us to study
the wave properties for the two different regimes.
Note, that through the whole paper we choose to discuss the
results in terms of the parameter $c_S^2/v_A^2$ instead of plasma
$\beta$. The former parameter is more relevant regarding wave
propagation. The ratio between these two parameters is equal to
$(c_S^2/v_A^2)/\beta = \gamma/2 =5/6$ (for an ideal gas). The
layers $c_S^2/v_A^2=1$ and $\beta=1$ are located very close in the
atmosphere. The maximum height separation between them is about 30
km.
The locations of the sources in both sets of simulations are
marked by yellow dots in Fig.~\ref{fig:spotmodel}. The duration of
each simulation run is about 50 minutes.

\subsection{Non-magnetic case}

\begin{figure*}
\centering
\includegraphics[width=16cm]{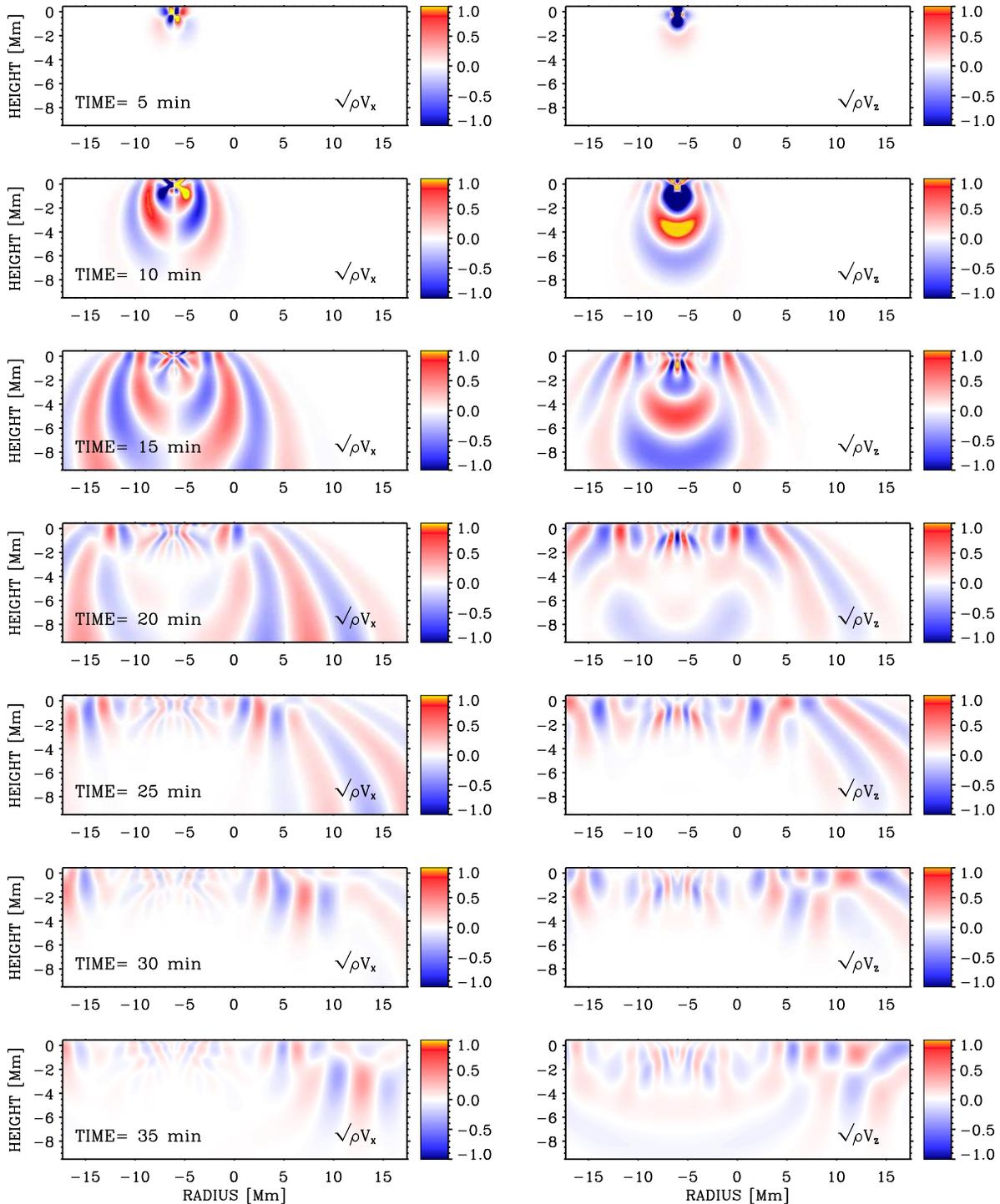}
\caption{Time series of snapshots of the horizontal (left panels)
and vertical (right panels) velocities in the non-magnetic
simulations with $X_0=$ 6 Mm. The velocities are scaled with
$\sqrt{\rho_0}$, the color scale is the same in both panels. The
negative heights correspond to sub-photospheric layers.}
\label{fig:nomagnetic}
\end{figure*}

Fig.~\ref{fig:nomagnetic} shows the time evolution of the
simulated horizontal and vertical velocities in the case of the
quiet Sun model. In the figure, the velocities are scaled with the
square root of the background density in order to make visible
variations in the deep layers.
The simulation reveals that the single acoustic source excites a
mixture of acoustic ($p$), gravity ($g$) and surface ($f$) modes,
each propagating with its own speed. The $p$ modes are the fastest
and run beyond the boundaries of the simulation domain after about
30 min. The slower $g$ and $f$ modes are concentrated in the upper
part of the simulation domain. It can be seen that the wave field
is symmetric with respect to the source location since the medium
is isotropic for the acoustic-gravity waves.

%%%%%%%%%%%%%%%%%%%%%%%%%%%%%%%%%%%%%%%%%%%%%%%%%%%%%%%%%%%%%%%%%%%%%%%%%%%%%%%%%%%%%%
%%%%%%%%%%%%%%%%%%%%%%%%%%%%%%%%%%%%%%%%%%%%%%%%%%%%%%%%%%%%%%%%%%%%%%%%%%%%%%%%%%%%%%%

\begin{figure}
\centering
\includegraphics[width=8cm]{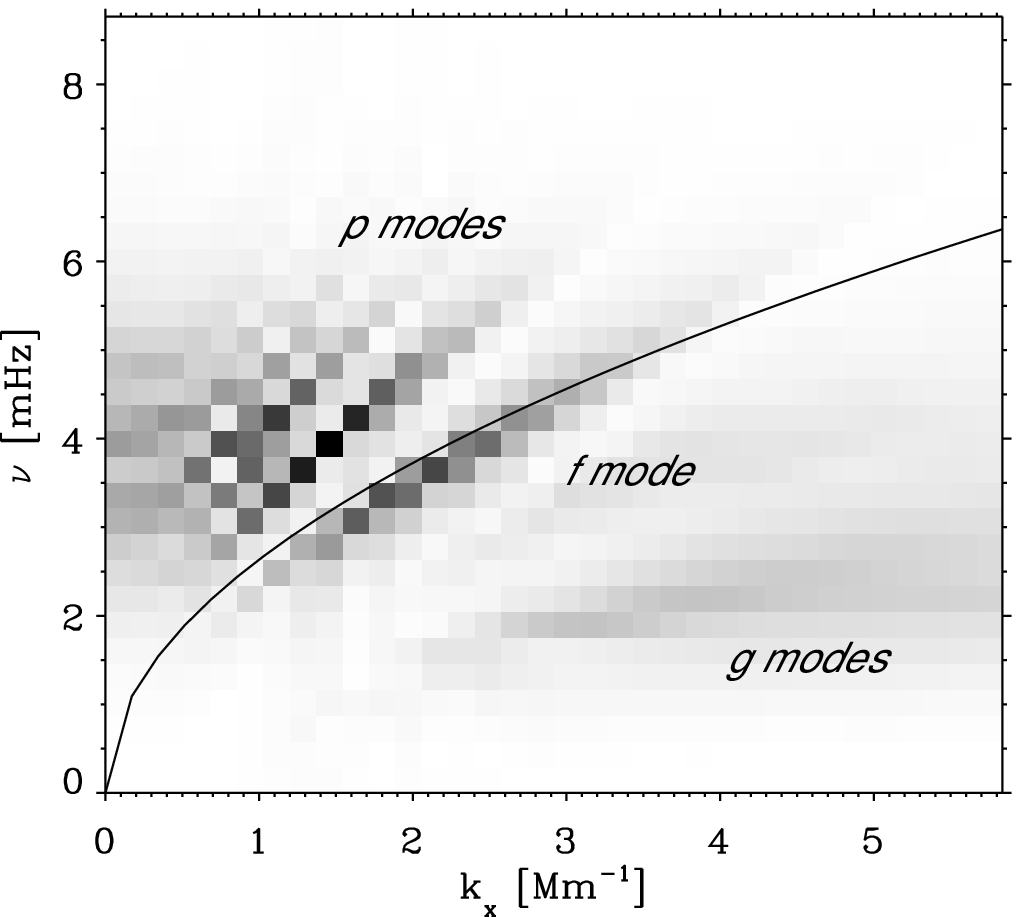}
\caption{The $k-\nu$ power spectrum of the photospheric vertical
velocity in the non-magnetic standard solar model simulation run.
Darker colors indicate larger power. The black line indicates the
dispersion relation of the $f$-mode: $\omega^2=g k_x$, where $g$
is the value of solar gravity acceleration at the surface. Several
$p$ modes can be distinguished above the $f$ mode. The $g$ modes,
located below the $f$ mode are also excited by the vertical force
source. These modes are present because the solar model has been
made convectively stable.} \label{fig:kw}
\end{figure}

The power spectrum of the photospheric vertical velocity ($k-\nu$
diagram) in the non-magnetic simulations is given in
Fig.~\ref{fig:kw}.  The diagram demonstrates the presence of
several ridges of acoustic ($p$-mode) oscillations. The darkest
bottom ridge corresponds to $f$ mode. The $f$-mode analytical
dispersion relation is also plotted in the image and shows a
reasonable agreement with the numerical results. The figure also
reveals the presence of a weak $g$-mode signal below the $f$-mode
ridge. These modes appear because the background model has been
made convectively stable.

%%%%%%%%%%%%%%%%%%%%%%%%%%%%%%%%%%%%%%%%%%%%%%%%%%%%%%%%%%%%%%%%%%%%%%%%%%%%%%%%%%%%%%
%%%%%%%%%%%%%%%%%%%%%%%%%%%%%%%%%%%%%%%%%%%%%%%%%%%%%%%%%%%%%%%%%%%%%%%%%%%%%%%%%%%%%%%
%%%%%%%%%%%%%%%%%%%%%%%%%%%%%%%%%%%%%%%%%%%%%%%%%%%%%%%%%%%%%%%%%%%%%%%%%%%%%%%%%%%%%%
\begin{figure*}
\centering
\includegraphics[width=16cm]{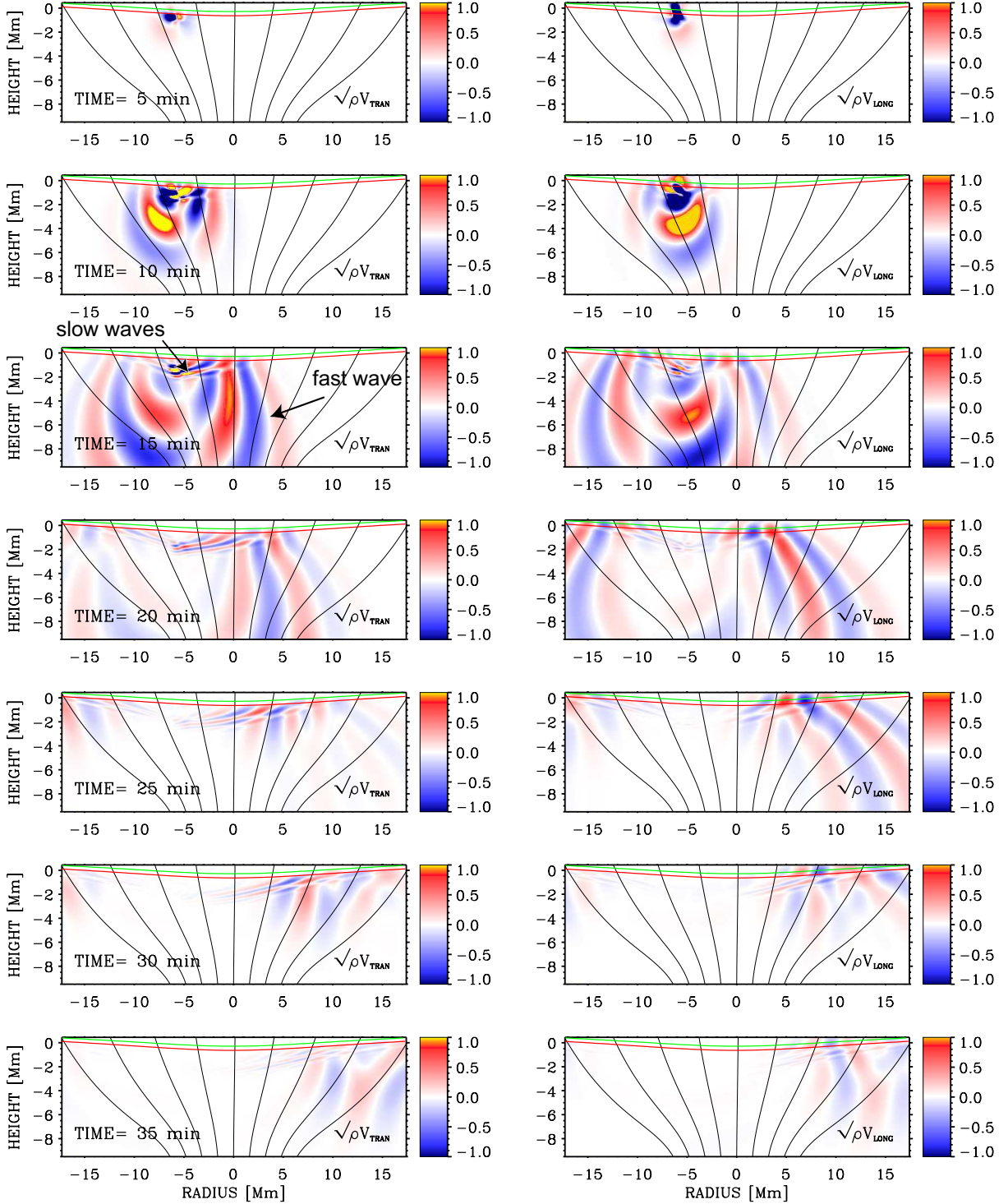}
\caption{Time series of snapshots of the transversal (left panels)
and longitudinal (right panels) velocities in the simulations with
$B_{\rm phot}$ =1.5 kG and $X_0=$ 6 Mm. The velocities are scaled
with the factor of $\sqrt{\rho_0}$. The color scales are the same
in all panels. The red and green lines are contours of constant
$c_S^2/v_A^2$, the red line corresponds to $v_A=c_S$, and the
green line corresponds to $c_S^2/v_A^2=0.1$. The black inclined
lines are the magnetic field lines. The negative heights
correspond to sub-photospheric layers. Note the presence of slow
waves, in addition to the modified $p$-modes (fast waves),
indicated by arrows in the time = 15 min panel.}
\label{fig:magnetic6}
\end{figure*}
%%%%%%%%%%%%%%%%%%%%%%%%%%%%%%%%%%%%%%%%%%%%%%%%%%%%%%%%%%%%%%%%%%%%%%%%%%%%%%%%%%%%%%%
%%%%%%%%%%%%%%%%%%%%%%%%%%%%%%%%%%%%%%%%%%%%%%%%%%%%%%%%%%%%%%%%%%%%%%%%%%%%%%%%%%%%%%
\begin{figure}
\centering
\includegraphics[width=8cm]{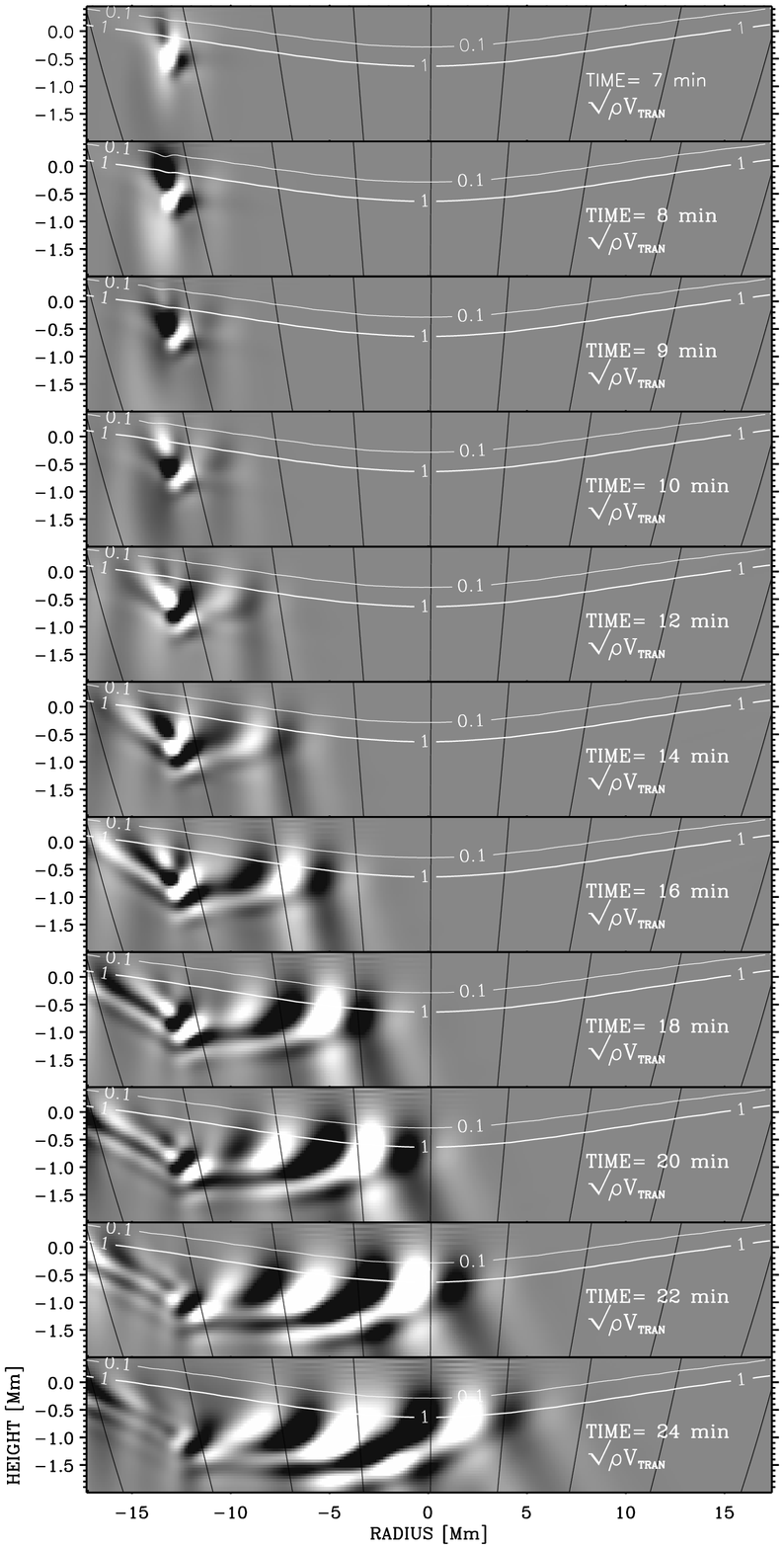}
\caption{Time series of snapshots of the transversal velocity in
the simulations with $B_{\rm phot}$ =1.5 kG and $X_0=$ 12 Mm. Only
the top part of the simulation box is shown and the vertical axis
has been expanded for better presentation. The velocities are
scaled with the factor of $\sqrt{\rho_0}$. Each panel is
normalized to its own maximum. } \label{fig:top12}
\end{figure}
%%%%%%%%%%%%%%%%%%%%%%%%%%%%%%%%%%%%%%%%%%%%%%%%%%%%%%%%%%%%%%%%%%%%%%%%%%%%%%%%%%%%%%%

\subsection{Waves in sunspot models}
%Only one example of a time series and description of the modes.

Figures~\ref{fig:magnetic6} and \ref{fig:top12} show the results
of the simulations in the sunspot model with $B_{\rm phot}$ = 1.5
kG and the source location at $X_0=$ 6 Mm and 12 Mm,
correspondingly. In the case of the simulations with magnetic
field, we use for illustrations the longitudinal (parallel to the
local field) and transversal (perpendicular to the field) velocity
components. This provides an effective separation of the fast and
slow magneto-acoustic waves in those regions of the simulation
domain where the characteristic propagation speeds are clearly
distinct ($v_A \ll c_S$ or $v_A \gg c_S$). Apart from a smaller
propagation speed, the slow mode is mainly magnetic in the region
where $v_A \ll c_S$, and has a displacement velocity that is
perpendicular to the magnetic field lines. This means that it
contributes mainly to the transversal velocity. On the other hand,
the fast mode is mainly acoustic, i.e., longitudinal, by its
nature. As this can be seen in Fig.~\ref{fig:magnetic6}, the
longitudinal component of the fast mode vanishes approximately
along the field line crossing the source. The direction of
propagation of the slow magneto-acoustic waves (defined as the
direction of their group velocity) is always parallel to the
magnetic field direction, irrespective of their phase velocity.
The direction of propagation of the fast magneto-acoustic waves is
independent of the field direction, and the group velocity is
parallel to the phase velocity.
Given these properties, the fast and slow modes can be
distinguished in the time evolution pictures of the transversal
and longitudinal velocity components, displayed in
Figures~\ref{fig:magnetic6} and \ref{fig:top12}.

In all simulations, the wave source is located in the region where
$c_S^2/v_A^2$ stay (roughly) between 0.1 and 10, and the field is
significantly inclined. In this region, both the pressure and
magnetic restoring forces are of the same order of magnitude.
Thus, the source generates a set of fast and slow magneto-acoustic
waves. The fast modes are an analog of the acoustic $p$-modes in
the non-magnetic simulations (see Fig. \ref{fig:nomagnetic}).
However, as it will be shown later in the paper, their behavior is
modified by the presence of the magnetic field. The slow modes did
not appear in the non-magnetic simulation.

Fig.~\ref{fig:magnetic6} shows that the fast wave propagating in
this complex magnetic field configuration causes variations in
both, longitudinal and transversal, components of the velocity of
the same order of magnitude. The initial wave front expands
propagating downwards. There is no symmetry in the wave field with
respect to the source location (unlike the case of the
non-magnetic simulation) since the wave propagation speed now
depends on the direction. In the deep layers, the dominant
velocity variations are due to the fast mode waves.
In the upper photospheric layers we observe mostly variations in
the longitudinal velocity (right panels in
Fig.~\ref{fig:magnetic6}). The transversal velocity variations,
caused by the fast magnetic waves ($c_S^2/v_A^2 \ll 1$), are
negligible in these layers.
The wave velocity field developed in the simulations with the
source located at $X_0=$ 6 and 12 Mm from the sunspot axis are
similar. The differences are quantitative, not qualitative in
nature.

Fig.~\ref{fig:top12} gives a more detailed view on the evolution
of the transversal velocity component in the upper part of the
simulation domain (note that the vertical axis has been expanded)
and allows to understand better the complex physics of the wave
propagation in the vicinity of the $c_S=v_A$ layer. An even better
view on the wave dynamics in this region can be gained by watching
the movie of the simulation, provided as on-line material to this
paper.
We observe in Fig.~\ref{fig:top12} the presence of the slow MHD
mode generated directly by the source. This mode propagates with a
visibly low speed downwards along the sunspot magnetic field
lines. It is localized close to the horizontal position  $X=-10$
Mm.

The fast waves generated by the source experience a significant
refraction and reflection due to the strong increase of the
Alfv\'en speed (see Fig.~\ref{fig:spot4000hz}) in the vicinity of
the $c_S=v_A$ layer (snapshots corresponding to times between 8
and 14 minutes in Fig.~\ref{fig:top12}).

In addition to the slow MHD waves generated directly by the source
there is another wave type.
It can be observed in Fig. \ref{fig:magnetic6} between 14 and 24
minutes of time as a perturbation with much smaller vertical
wavelength compared to that of the fast modes in the sunspot
interior.
We can observe from Fig. \ref{fig:magnetic6} that the propagation
speed of this low-wavelength disturbance is comparable to that of
the fast modes.
Unlike the slow MHD waves, these waves propagate horizontally
across the sunspot.
Thus, these waves are different from either fast or slow MHD waves
in several aspects: (1) they have a vertical wavelength that is
much smaller than that of the fast waves; (2) they propagate with
speeds similar to the fast waves (note, however, that they
propagate in the region where the sound speed is slightly larger
than the Alfv\'en speed); (3) they propagate across the field, a
behavior impossible for slow waves in this region; (4) the waves
that posses similar characteristics (superficial horizontal
propagation, smaller wavelength, etc) in non-magnetic simulations
are the $f$-mode waves. From these arguments we suggest that we
see a wave type with properties that are a mixture of magnetic and
$f$-mode waves. We call them ``surface magneto-gravity'' waves.
These waves were mentioned previously by, \eg\
\citet{Schwartz+Stein1975}, who described their properties in an
isothermal atmosphere.
The variations produced by the surface magneto-gravity waves
decrease rapidly with depth and disappear almost completely below
$-3$ Mm.

%%%%%%%%%%%%%%%%%%%%%%%%%%%%%%%%%%%%%%%%%%%%%%%%%%%%%%%%%%%%%%%%%%%%%%%%%%%%%%%%%%%%%%
\begin{figure*}
\centering
\includegraphics[width=5.7cm]{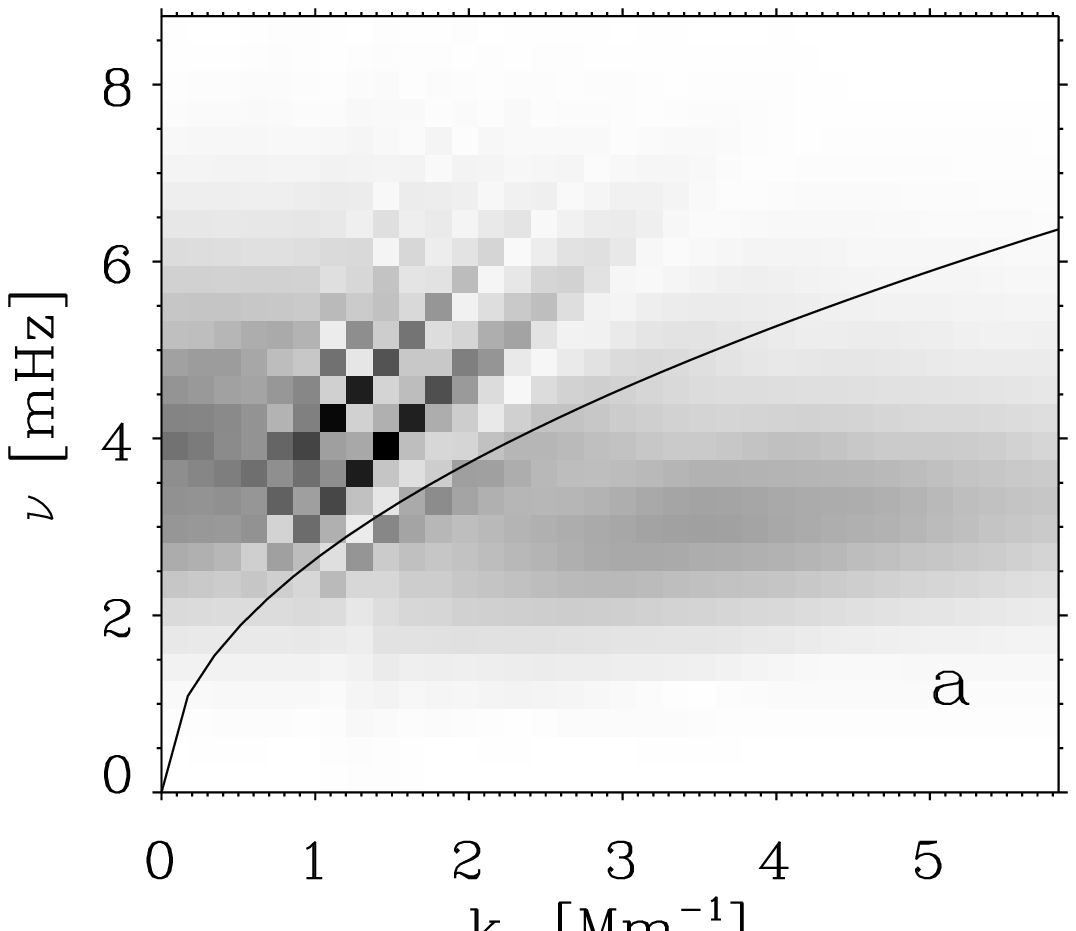}
\includegraphics[width=5.7cm]{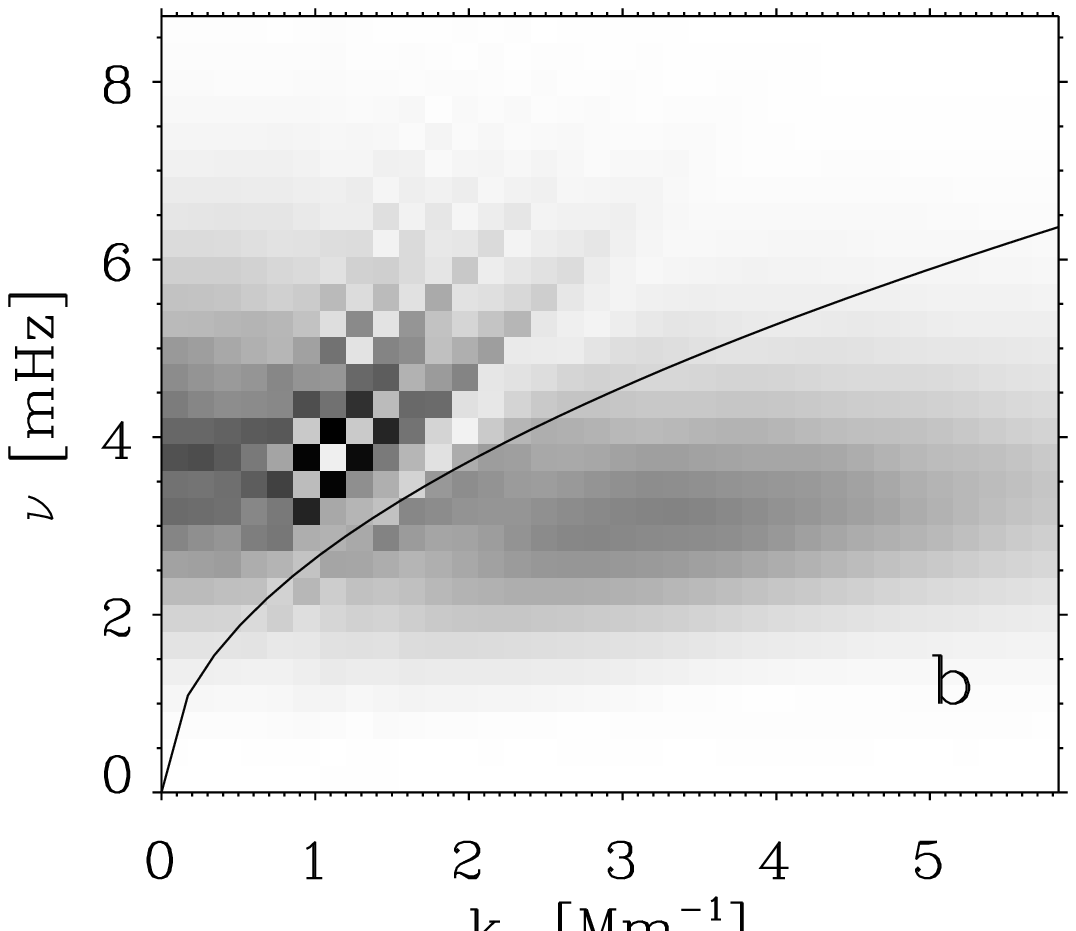}
\includegraphics[width=5.7cm]{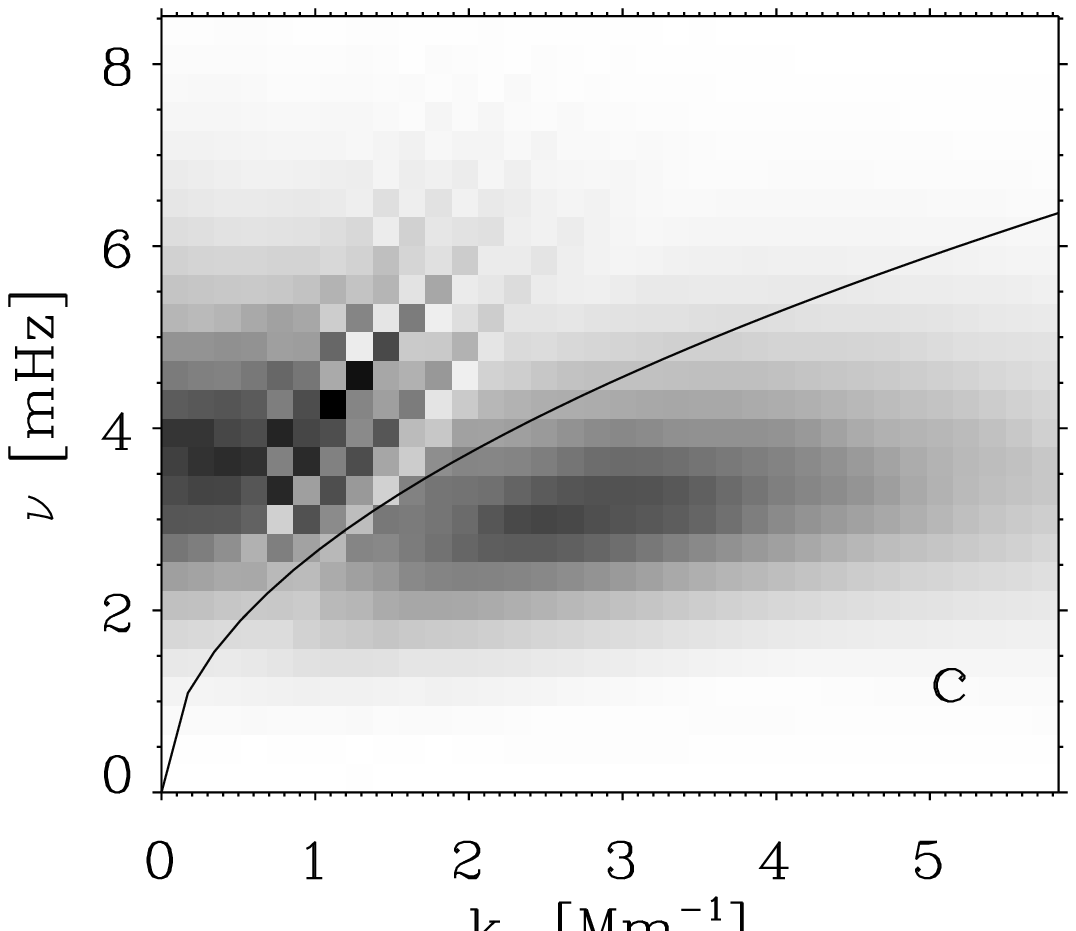}
\caption{The $k-\nu$ power spectra of the photospheric vertical
velocity in the simulation runs for the source position $X_0$= 12
Mm and $B_{\rm phot}$=0.9 (a), 1.5 (b) and 2.4 (c) kG. Darker
colors indicate larger power. Note the change of the power spectra
with the increasing field strength. } \label{fig:kw3}
\end{figure*}
%%%%%%%%%%%%%%%%%%%%%%%%%%%%%%%%%%%%%%%%%%%%%%%%%%%%%%%%%%%%%%%%%%%%%%%%%%%%%%%%%%%%%%%

The power spectra of the photospheric vertical velocity in the
magnetic simulations for different $B_{\rm phot}$ and the source
location at 12 Mm from the sunspot axis are compared in
Fig.~\ref{fig:kw3}. Figure~\ref{fig:kw3} shows that for the
magnetic models the $k-\nu$ diagrams become progressively more and
more different from the non-magnetic $k-\nu$ diagram
(Fig.~\ref{fig:kw}), with increasing magnetic field strength. Even
in the case of the weakest field, $B_{\rm phot}$ = 0.9 kG, the
inclination of the $p$-mode ridges is larger than in the pure
acoustic case. The inclination of the ridges increases with
increasing magnetic field strength. The dispersion relation for
the $p$-mode ridges depends on the reflection conditions for waves
in the upper layers. In the absence of magnetic field, the waves
with different frequencies reflect at different heights in the
photosphere and below due to the rapid change of the cut-off
frequency with height. When the magnetic field is introduced, the
upper reflection condition changes even if the field is as weak as
in the $B_{\rm phot}$ = 0.9 kG model. This happens because the
propagation path of the high frequency fast waves is modified
strongly by the rapid increase of the Alfv\'en speed since these
waves penetrate higher up in the atmosphere. The reflection height
for such waves becomes lower compared to the non-magnetic case.
This naturally produces the change of the dispersion relation
observed at the $k-\nu$ diagrams in Fig.~\ref{fig:kw3}.

%The presence of the weak $f$-mode can be distinguished in the
%simulation with $B_{\rm phot}$ = 0.9 kG. This mode disappears for
%larger field strengths and is not present in the simulation with
%$B_{\rm phot}$ = 2.4 kG. In addition to the modified $p$-modes,
%the magnetic $k-\nu$ diagrams have power in the region of
%$g$-modes. {\bf This power corresponds now the to slow MHD modes
%and magneto-gravity modes. As was observed in the time evolution
%of the velocities in simulations (Figs.~\ref{fig:magnetic6} and
%\ref{fig:top12}), the slow  and magneto-gravity waves have much
%smaller wavelength for the same frequency. Thus, they are located
%at the $k-\nu$ diagrams in the region of large horizontal
%wavenumbers $k_x$. } The contribution of the slow-mode power
%increases with increasing field strength.

%%%%%%%%%%%%%%%%%%%%%%%%%%%%%%%%%%%%%%%%%%%%%%%%%%%%%%%%%%%%%%%%%%%%%%%%%%%%%%%%%%%%%%
\begin{figure}
\centering
\includegraphics[width=9cm]{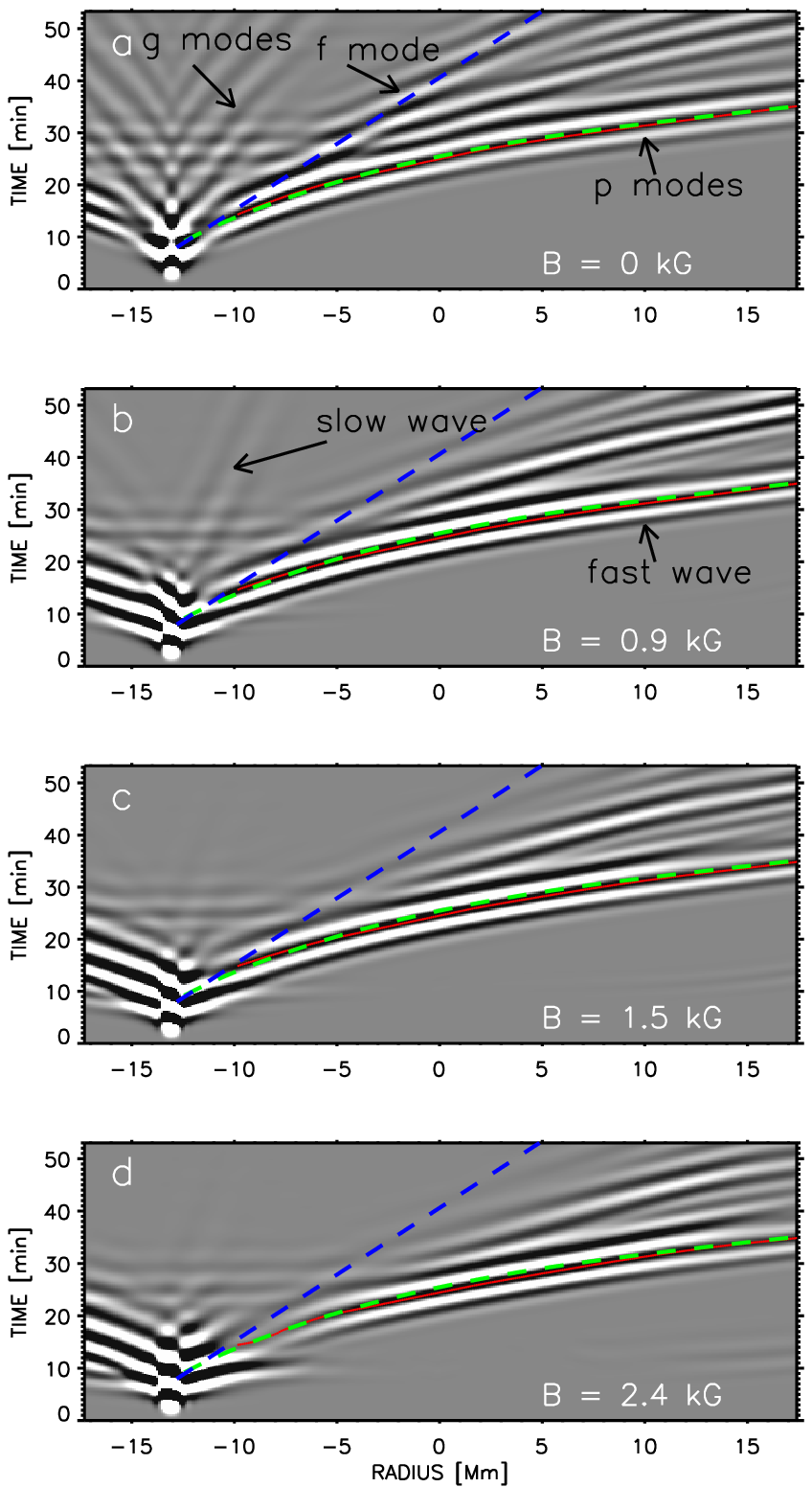}
\caption{The time-distance diagrams of the  photospheric vertical
velocity for the simulations with (a) no magnetic field; (b)
sunspot model with $B_{\rm phot}$ = 0.9 kG; (c) sunspot model with
$B_{\rm phot}$ = 1.5 kG; (d) sunspot model with $B_{\rm phot}$ =
2.4 kG. The source is located at $X_0=$ 12 Mm on the left of the
sunspot axis. The red curves are the phase travel times calculated
using the Gabor wavelet fit to the time-distance curves (see the
text). The green dashed curves are theoretical time-distance
curves for $p$-modes and the standard solar model S in the absence
of the magnetic field. The blue dashed lines are the theoretical
time-distance curves for the $f$ mode calculated as
$X=(1/2)(g/\omega) t$ for the central frequency $\omega=2\pi/300$
of the wave source. } \label{fig:td}
\end{figure}
%%%%%%%%%%%%%%%%%%%%%%%%%%%%%%%%%%%%%%%%%%%%%%%%%%%%%%%%%%%%%%%%%%%%%%%%%%%%%%%%%%%%%%%
%%%%%%%%%%%%%%%%%%%%%%%%%%%%%%%%%%%%%%%%%%%%%%%%%%%%%%%%%%%%%%%%%%%%%%%%%%%%%%%%%%%%%%
\begin{figure}
\centering
\includegraphics[width=9cm]{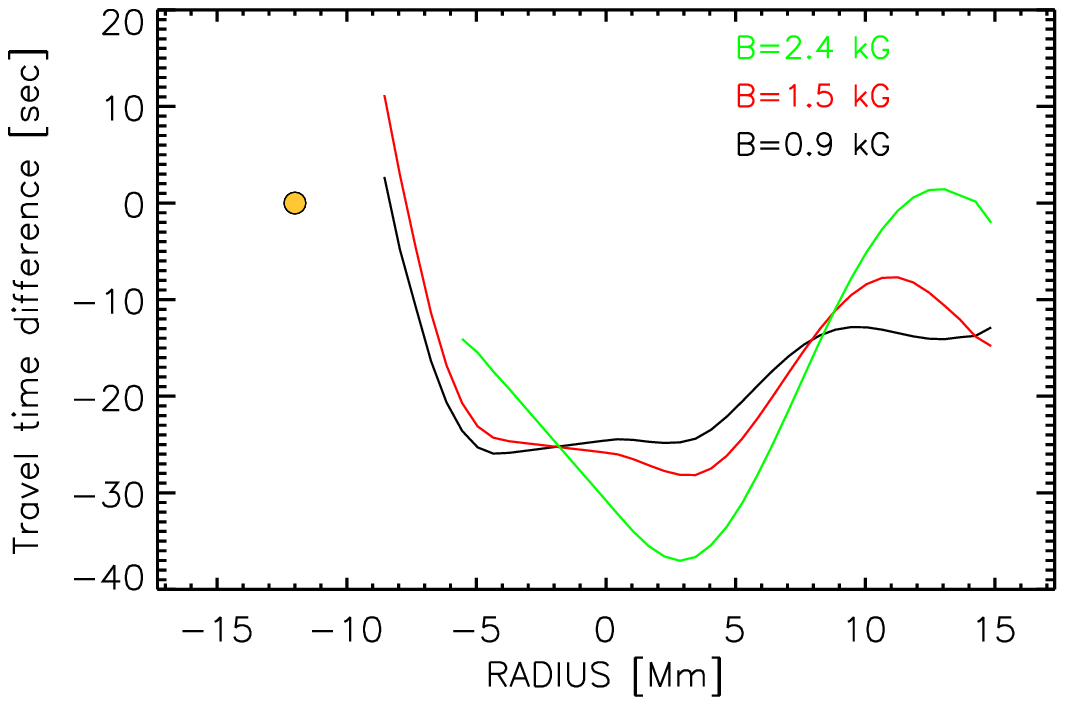}
\caption{The phase travel-time difference calculated between the
travel times measured in simulations without magnetic field and
the simulations with different model sunspots (indicated in the
figure) as a function of horizontal distance. The source location
at $X_0=$ 12 Mm left to the sunspot axis is marked by yellow
circle.} \label{fig:traveltime}
\end{figure}
%%%%%%%%%%%%%%%%%%%%%%%%%%%%%%%%%%%%%%%%%%%%%%%%%%%%%%%%%%%%%%%%%%%%%%%%%%%%%%%%%%%%%%%

\subsection{Time-distance diagrams and travel time estimates}
% aqui

In order to quantify the effects of magnetic field on the travel
time measurements of local helioseismology, we have constructed
time-distance diagrams for the vertical velocity. Such diagrams
were constructed for the photospheric velocity measured at the
optical depth log$\tau_5=-1.6$ (see Sect. 3) in the sunspot
models. Since in our simulations the waves are excited by a single
source, there is no need to compute the velocity correlations
between the different surface points. Thus, the time-distance wave
propagation diagrams are constructed by plotting the velocity
variations as a function of time for all horizontal points at
constant log$\tau_5$ level.
Fig.~\ref{fig:td} shows the time-distance diagrams in the
non-magnetic simulations and the simulations with progressively
increasing magnetic field strength.

In the time-distance diagram for the non-magnetic case
(Fig.~\ref{fig:td}, a), the two lowest system of ridges are due to
the $p$ modes. The green dashed line in Fig.~\ref{fig:td}a gives a
theoretical dispersion relation for these modes, calculated for
the standard solar model S.
Above them, the third system of ridges comes from the $f$ mode
waves. It can be seen from the inclination of the ridges that the
phase and group speeds of this mode are different from those of
the $p$ modes, confirming that they belong to the $f$-mode. The
theoretical time-distance curve for the $f$-mode (blue dashed line
in Fig.~\ref{fig:td}a) is in a good agreement with the numerical
results.
The system of ridges at the top part of the diagram above the
source corresponds to the $g$ modes. These modes propagate with
much lower speed than the $p$ and $f$ modes, in agreement with
their appearance at the $k-\nu$ diagram in Fig.~\ref{fig:kw}.

The time-distance diagrams in the simulations with magnetic field
have different properties (Fig.~\ref{fig:td}, b, c, and d).
$p$-mode-like ridges are present in all the cases.
As in the non-magnetic case, we have added theoretical curves for
$p$-modes and the $f$-mode in the absence of the magnetic field
(green and blue dashed curves, in Fig.~\ref{fig:td}b, c, and d).
These curves help identifying the fast modes as the analog of the
non-magnetic $p$-modes propagating with a speed modified by the
magnetic field.
The $f$- and $g$-mode ridges do not appear, in agreement with the
$k-\nu$ diagram in Fig.~\ref{fig:kw3}. Instead, weak traces of the
slow waves are present in the top part of the diagram. This mode
is not symmetric with respect to the source location and has only
surface signatures on the side of the source where the magnetic
field is larger. At the log$\tau_5=-1.6$, the slow mode is mostly
prominent in the sunspot simulation with weak field $B_{\rm phot}
=$ 0.9 kG. It becomes weaker but still can be distinguished in the
models with stronger field. The origin of this mode is different
from the slow modes generated by the source in the sunspot
interior (Fig.~\ref{fig:top12}) and needs further study.

At first glance, the $p$-mode ridges on the time-distance diagrams
seem not to be affected much by the magnetic field. A more
detailed investigation shows that the $p$-mode ridge, at the
bottom,  becomes thicker, \ie\ the relative contribution of the
different frequencies in the wave packet changes with magnetic
field.
We have performed a fit to the time-distance diagrams in order to
find the phase and group travel times of waves in each case.
At each horizontal position starting from 3 Mm from the source we
fit the time velocity curve with a Gabor's wavelet:
\begin{equation}
G(\tau)=A\cos[\omega_0(\tau-\tau_p)]\exp\left[
-\frac{\delta\omega^2}{4}(\tau-\tau_g)^2\right],
\end{equation}
where $A$ is the amplitude, $\omega_0$ is the central frequency,
$\tau_p$ and $\tau_g$ are the phase and group travel times
respectively, and $\delta\omega$ is the bandwidth
\citep{Kosovichev+Duvall1997}. The fit has four free parameters:
$\omega_0$, $\tau_p$, $\tau_g$, and $\delta\omega$. The red curves
plotted in the diagrams of Fig.~\ref{fig:td} are the phase travel
times $\tau_p$ obtained after the fit.

Fig.~\ref{fig:traveltime} shows the differences in the phase
travel-times between the magnetic and non-magnetic simulations.
Negative values mean that waves in the magnetic simulations
propagate faster. The travel-time difference curves have a complex
dependence on the distance.
The difference between the magnetic and non-magnetic cases is
larger at locations closer to the sunspot axis, where the magnetic
field is stronger, and getting smaller at $15$ Mm far from the
axis. This happens because the waves with larger skip distances
travel deeper inside the sunspot and are less affected by the
magnetic surface effects.
The travel time difference depends on the magnetic field strength
of the model. The model with the smallest field $B_{\rm phot}$ =
0.9 kG gives a maximum difference in the travel times of about 25
seconds, while in the model with the largest field strength
$B_{\rm phot}$ = 2.4 kG, this difference becomes about $35$
seconds.
The results presented in Figs.~\ref{fig:td} and
\ref{fig:traveltime} are calculated for the photospheric vertical
velocities taken at constant log$\tau_5$ level. These results do
not change much if we take velocities at a constant geometrical
height $Z=100$ km, instead of constant log$\tau_{5}$=$-1$.6. Only
a minor additional phase shift is introduced in the travel time
variations in Fig.~\ref{fig:traveltime} making them about 5
seconds smaller close to the sunspot axis.

%%%%%%%%%%%%%%%%%%%%%%%%%%%%%%%%%%%%%%%%%%%%%%%%%%%%%%%%%%%%%%%%%%%%%%%%%%%%%%%%%%%%%%
\begin{figure}
\centering
\includegraphics[width=8cm]{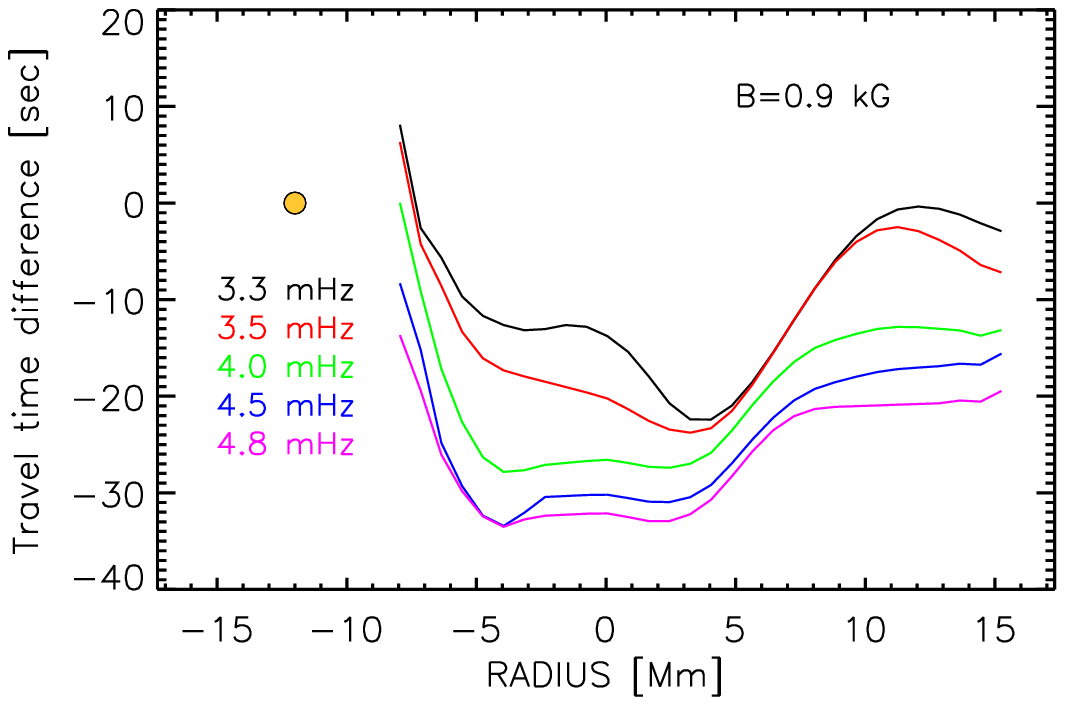}
\includegraphics[width=8cm]{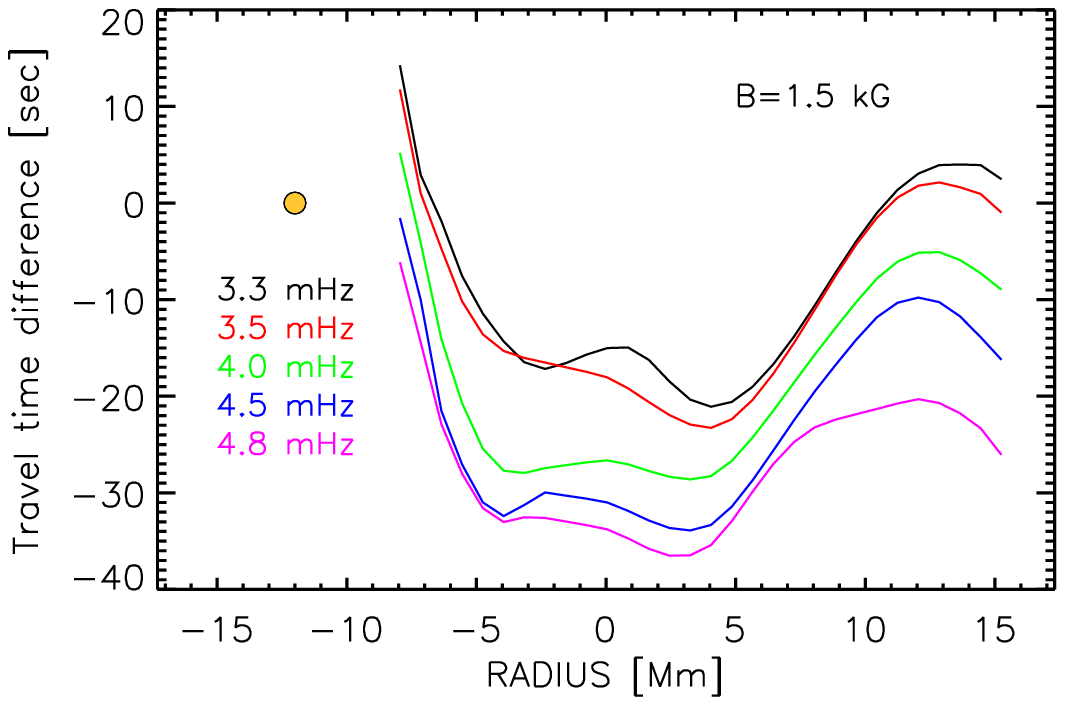}
\includegraphics[width=8cm]{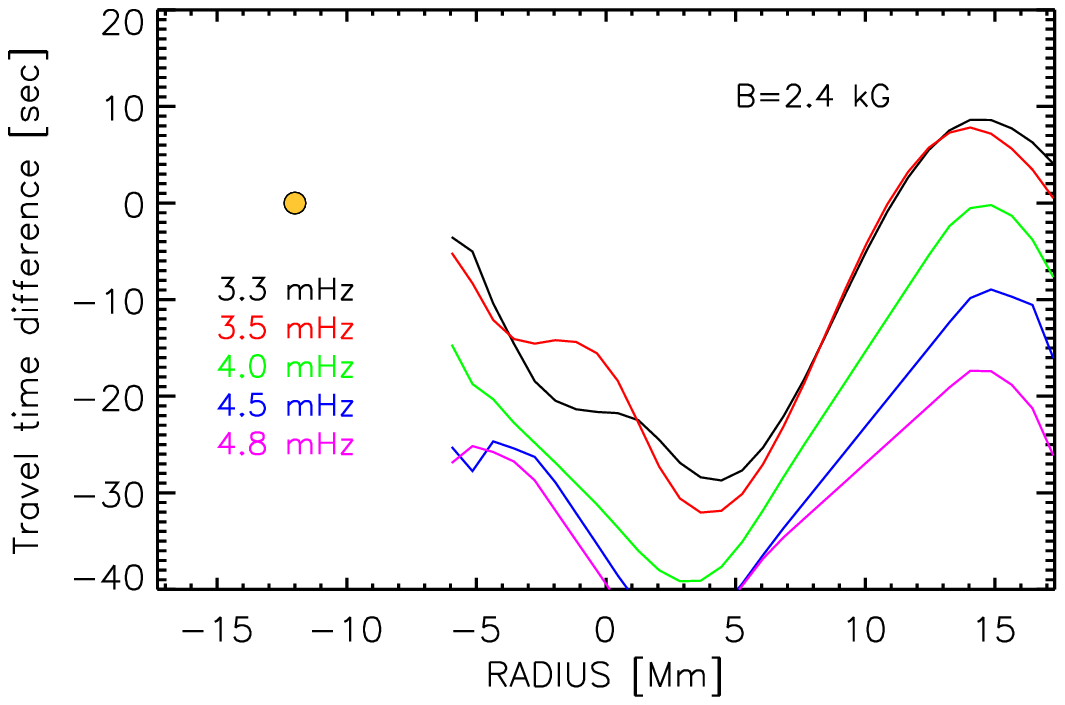}
\caption{The phase travel time difference between the magnetic and
non-magnetic cases as a function of the horizontal distance,
calculated for different frequency filtering of the simulations
with the source located at 12 Mm. The different colors correspond
to different wave frequencies. Calculations done for the model
sunspot with $B_{\rm phot}$ = 0.9 kG (top), 1.5 kG (middle) and
2.4 kG (top).} \label{fig:frequency}
\end{figure}
%%%%%%%%%%%%%%%%%%%%%%%%%%%%%%%%%%%%%%%%%%%%%%%%%%%%%%%%%%%%%%%%%%%%%%%%%%%%%%%%%%%%%%%

% discussion on the frequency dependence

The travel-time differences discussed above were calculated for
the central frequency of the wave packet. However, there has been
a discussion in the recent literature regarding the frequency
dependence of the travel time measurements
\citep{Couvidat+Birch+Kosovichev2006, Couvidat+Rajaguru2007,
Rajaguru2008, Braun+Birch2008}. It has been obtained that, if the
frequency filtering is applied to the data, in addition to the
phase speed filtering, the resulting travel-time differences show
a strong frequency dependence. High frequency waves have, in
general, larger travel-time differences \citep{Rajaguru2008}. In
order to investigate this effect in our simulations, we filtered
the simulated photospheric vertical velocities in the five
frequency ranges: 3.3, 3.5, 4.0, 4.5 and 4.8 mHz. We applied a
Gaussian filter with a FWHM of 1 mHz. The resulting travel-time
differences are shown in Fig.~\ref{fig:frequency}, for all three
model sunspots.

We find that the travel-time differences for each individual
frequency behave similarly to the  travel-time differences
calculated without filtering. The maximum difference is achieved
for distances closer to the sunspot axis ($\pm 5$ Mm), decreasing
for the waves with large skip distances. A strong frequency
dependence is evident from Fig.~\ref{fig:frequency}. Waves with
higher frequencies have larger travel-time differences. It means
that the propagation speed of waves increases with frequency. The
magnitude of the frequency dependence that we obtain from the
simulations is in agreement with the observations of
\citet{Rajaguru2008}. In that paper travel-time differences change
from, about, $-15$ to $-30$ seconds between 3 and 5 mHz, similar
to our results in Fig.~\ref{fig:frequency}.

Interestingly, the travel-time differences get slightly positive
for the distances within first $5-6$ Mm from the source location
and for lower frequencies between 3.3 and 4.0 mHz. This effect
needs further investigation.
%Possibly, the change of the sign of
%the travel-time differences is a temperature effect. The
%temperature is lower inside the sunspot and the speed of the waves
%in the surface layers is lower due to the temperature effects.
%{\bf In our opinion, the change of the sign of the travel-time
%differences is a temperature effect. Since the temperature is
%lower inside the sunspot, the speed of the waves in the surface
%layers is also lower.} Qualitatively this corresponds to the
%observations of \citet{Kosovichev+etal2000}.

%%%%%%%%%%%%%%%%%%%%%%%%%%%%%%%%%%%%%%%%%%%%%%%%%%%%%%%%%%%%%%%%%%%%%%%%%%%%%%%%%%%%%%
\begin{figure*}
\centering
\includegraphics[width=16cm]{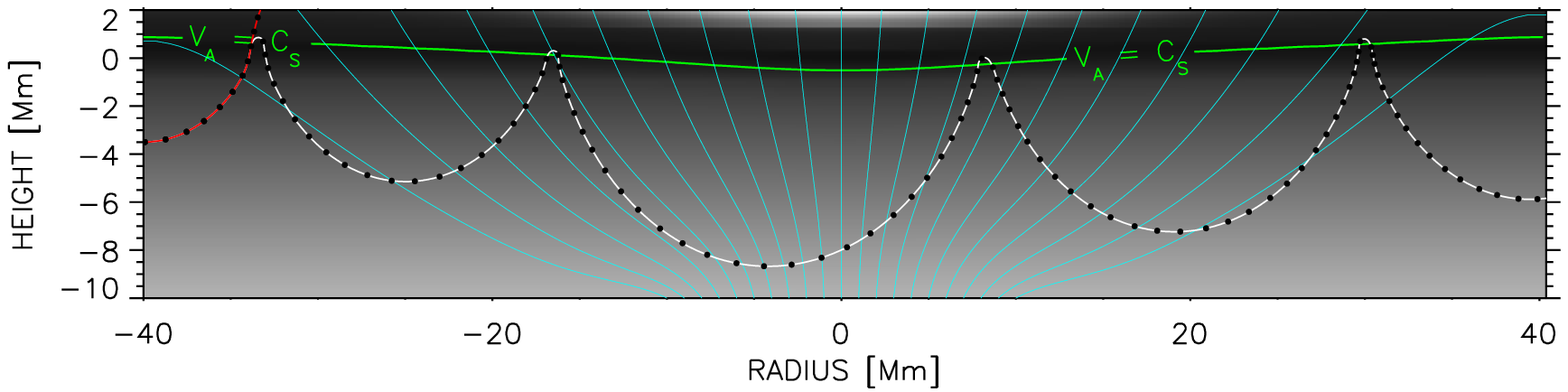}
\includegraphics[width=16cm]{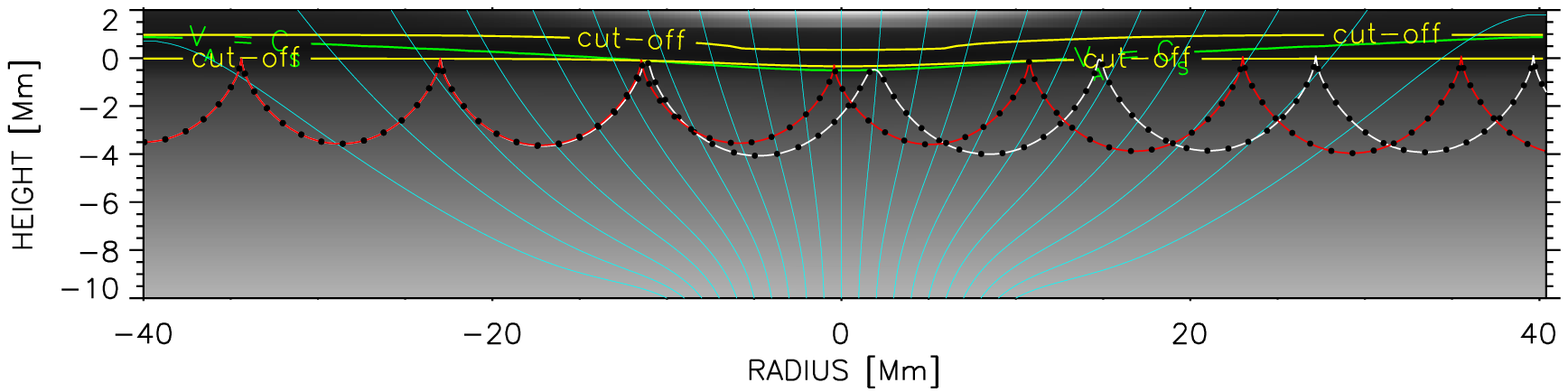}
\includegraphics[width=16cm]{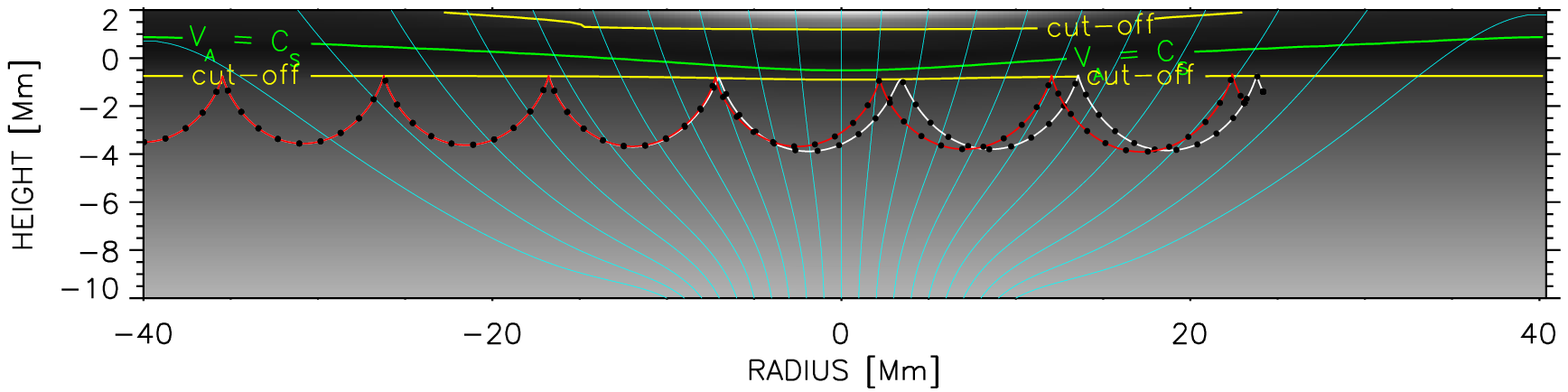}
\caption{The wave paths of the fast mode waves launched from the
same lower turning point below the photosphere propagating through
the model sunspot with $B_{\rm phot}$ = 2.4 kG for different
frequencies: 6.6 mHz (top); 4.5 mHz (middle) and 3 mHz (bottom).
The white curves are the wave paths when the magnetic field is
taken into account. The red curves are the wave paths in the
sunspot models with the same thermal properties, but setting
$B=0$. The black dots on the trajectory are separated from each
other by 1 minute in time. The green curve marks the position of
the $v_A=c_S$ layer. The yellow curves mark the heights where the
frequency of the wave is equal to the local acoustic cut-off
frequency. The background grey image is acoustic speed $c_S$.}
\label{fig:wkbpath}
\end{figure*}
%%%%%%%%%%%%%%%%%%%%%%%%%%%%%%%%%%%%%%%%%%%%%%%%%%%%%%%%%%%%%%%%%%%%%%%%%%%%%%%%%%%%%%%

%%%%%%%%%%%%%%%%%%%%%%%%%%%%%%%%%%%%%%%%%%%%%%%%%%%%%%%%%%%%%%%%%%%%%%%%%%%%%%%%%%%%%%
\begin{figure}
\centering
\includegraphics[width=8cm]{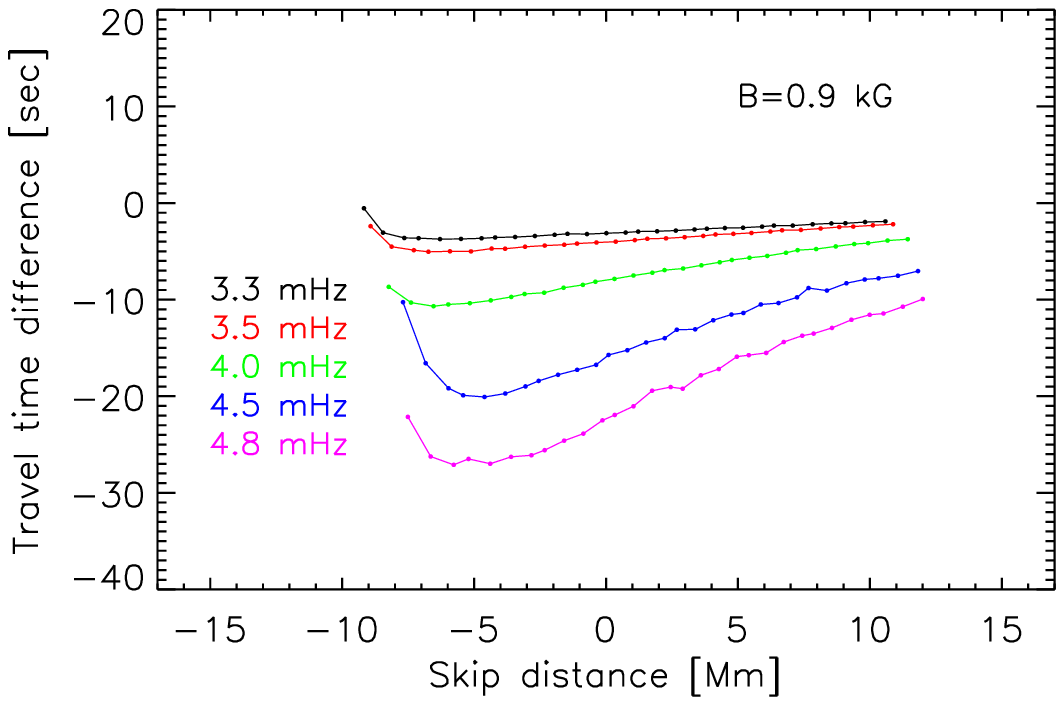}
\includegraphics[width=8cm]{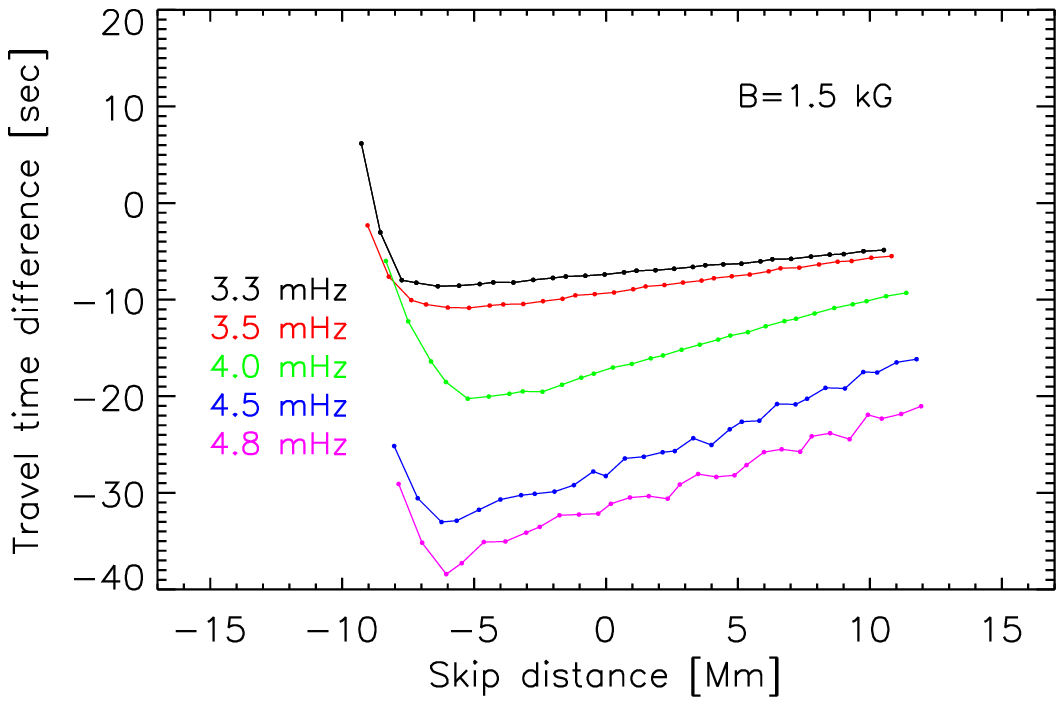}
\includegraphics[width=8cm]{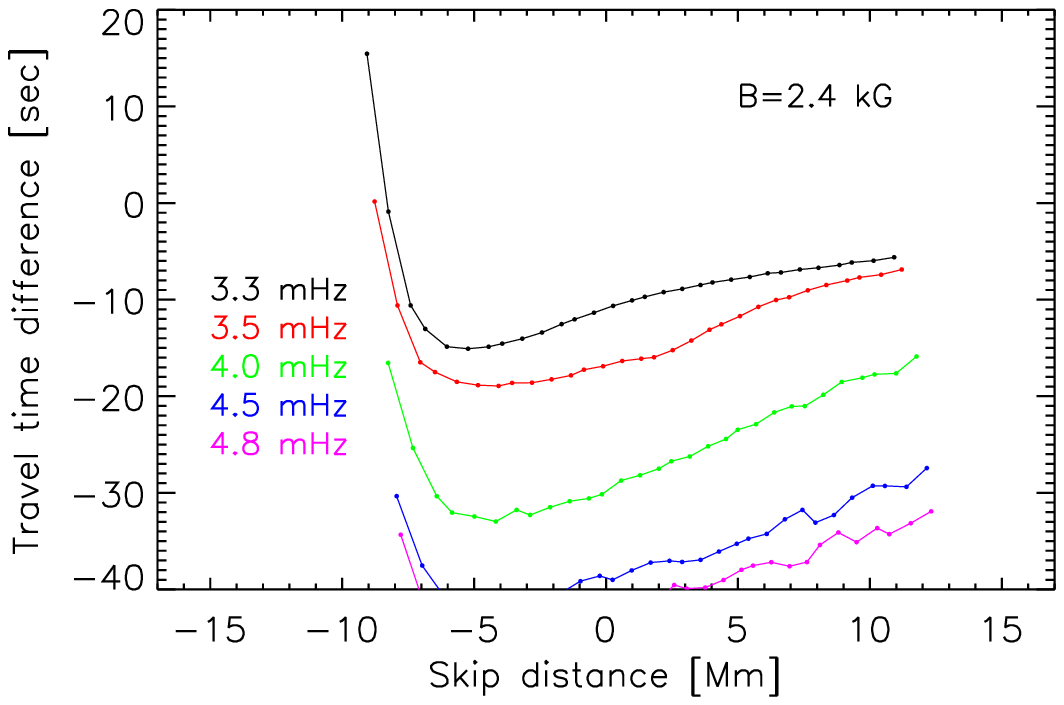}
\caption{The travel time difference between magnetic and
non-magnetic cases as a function of the skip distance, calculated
in the eikonal approximation for the annuli size of 12 Mm. The
lines of different colors correspond to different wave
frequencies. The calculations are done for the sunspot models with
$B_{\rm phot}$ = 0.9 kG (top), 1.5 kG (middle) and 2.4 kG (top).}
\label{fig:wkbtt}
\end{figure}
%%%%%%%%%%%%%%%%%%%%%%%%%%%%%%%%%%%%%%%%%%%%%%%%%%%%%%%%%%%%%%%%%%%%%%%%%%%%%%%%%%%%%%%

\section{Eikonal approximation solutions}

In order to understand more deeply the numerical results, we
present here an approximate solution of the governing MHD
equations for small perturbations in the eikonal approximation
\citep[see recent applications in \eg\ ][]{McLaughlin+Hood2006,
McLaughlin+etal2008, Khomenko+Collados2006}.
The eikonal approach assumes that the wavelength of the
perturbation is much smaller than the characteristic scale of the
variations of the background model \citep{Gough2007}.
In the zero-order eikonal approximation, we neglect the variation
of the wave amplitude and consider only the variation of its
phase, i.e., we assume that all the variables in
(\ref{eq:den}--\ref{eq:ind}) depend on $x, z$ and time as  $U=a
\cdot e^{i \phi(x,z)}\cdot e^{-i\omega t}$ (where $a$ is
constant).
We make use of the recent developments of \citet{Cally2006} and
\citet{Moradi+Cally2008} that allow to incorporate the effects of
the acoustic cut-off frequency in the eikonal solution for MHD
waves.
In this approximation after some mathematical simplifications, the
following first-order non-linear partial differential equation is
obtained:
\begin{eqnarray}
F(x,z,\phi,p,q) & = & \omega^4  - \omega^2(c_S^2 + v_A^2)(p^2 +
q^2) \\ \nonumber &  + &
 c_S^2(p^2 + q^2)(v_{Ax}p + v_{Az}q)^2 \\ \nonumber & - & \omega_c^2(\omega^2 -
c_S^2q^2) + c_S^2N^2p^2 = 0 \,, \label{eq:disp}
\end{eqnarray}
where $p=\frac{\partial \phi}{\partial x}$ and $q=\frac{\partial
\phi}{\partial z}$. Parameters $p$ and $q$ are equivalent to the
horizontal and vertical wave numbers, respectively. Parameter $N$
is the Brunt-V\"ais\"al\"a frequency, $N^2=g/H - g^2/c_S^2$. In
the following calculations, we use an expression for the cut-off
frequency for an isothermal atmosphere $\omega_c = c_S/2H$.
Equation (\ref{eq:disp}) is solved by  Charpit's method of
characteristics by transforming it into the following system of
ordinary differential equations:
\begin{eqnarray}
\label{eq:char1}
\frac{dp}{ds} & = & - \frac{\partial F}{\partial x} \\
\label{eq:char2}
\frac{dq}{ds} & = & - \frac{\partial F}{\partial z} \\
\label{eq:char3}
\frac{dx}{ds} & = &  \frac{\partial F}{\partial p} \\
\label{eq:char4}
\frac{dz}{ds} & = &  \frac{\partial F}{\partial q}
\end{eqnarray}
Here, variable $s$ is the distance along the characteristic wave
propagation path. These four ODEs are solved numerically using a
fourth-order Runge-Kutta method. We investigate only fast-mode
solutions of the equations. Parameters $c_S$, $v_A$, $N$ and
$\omega_c$ are taken from the sunspot models.
The solution of Eqs.(\ref{eq:char1}--\ref{eq:char4}) gives us
$x(s)$, $z(s)$, $p(s)$ and $q(s)$ along the wave path $s$. The
lines $x(z)$ give the trajectory of the group velocity of the
solution. The phase along the wave path can be calculated as
\citep{Moradi+Cally2008}:
\begin{equation}
S(\vec{r})=\int{\vec{k}d\vec{r}} - \omega t
\end{equation}
where $\vec{r}=(x,z)$ and $\vec{k}=(p,q)$. The phase function
$S(\vec{r})$ divided by the wave frequency $\omega$ gives us an
estimate of the phase travel time.

First, we study the paths through the sunspot models of fast
magneto-acoustic waves of a given frequency $\omega$. Waves are
launched from the same lower turning point at some depth below the
surface. Fig.~\ref{fig:wkbpath} gives the examples of solutions
$x(z)$ for the fast wave paths for the frequencies 6.6 (top), 4.5
(middle) and 3.0 (bottom) mHz. As initial condition for the
solution of the equations (\ref{eq:char1}--\ref{eq:char4}) we take
in all the cases $x=-40$ Mm, $z=-3.5$ Mm and $q=0$. The initial
value of $p$ is found from the approximated solution of the
dispersion relation (\ref{eq:disp}) at $q=0$.

% aqui

At each frequency, we calculate the wave paths for two cases. In
the first case (red lines in Fig.~\ref{fig:wkbpath}) we set the
Alfv\'en speed $v_A$ to zero (\ie\ eliminate the magnetic field),
but keep the sound speed structure and frequencies $N$ and
$\omega_c$ from the sunspot model. In the second case, we
calculate the wave path in the sunspot model taking into account
magnetic field terms (white lines in Fig.~\ref{fig:wkbpath}). As
can be observed in Fig.~\ref{fig:wkbpath}, the paths of the waves
in the magnetic and non-magnetic cases are significantly
different. Black dots on the curves mark the positions separated 1
minute in time in order to outline the quantitative differences in
phase times spent by the waves on their paths.

The frequency of the wave at the top panel of
Fig.~\ref{fig:wkbpath} (6.6 mHz) is larger than the atmospheric
cut-off. Thus, in the non-magnetic case, this wave (red line)
propagates to the upper atmospheric layers without reflection and
goes outside the simulation domain. This does not happen in the
magnetic case. The green horizontal line in Fig.~\ref{fig:wkbpath}
marks the position of the layer where $c_S = v_A$. As the fast
magneto-acoustic wave reaches this layer from below, its
propagation properties become progressively more affected by the
magnetic field. The fast magneto-acoustic wave in the region $v_A
> c_S$ experiences the refraction due to the rapid increase of the $v_A$ with height
\citep[a clear example of such behavior can be seen in the
simulations presented by][]{Khomenko+Collados2006}. Finally the
fast magneto-acoustic wave reflects back to the deep layers at
some height above the $c_S = v_A$. In the deep layers, this wave
experiences again the refraction and reflection, but now due to
the rapid increase of sound speed $c_S$.  Thus, in this model, the
high-frequency fast waves in the magnetic case are trapped below
the photosphere due to magnetic effects.

If we reduce the wave frequency to 4.5 mHz (middle panel in
Fig.~\ref{fig:wkbpath}), then in both the magnetic and
non-magnetic cases the wave paths have an upper reflection point.
The height where the wave frequency is equal to the atmospheric
cut-off frequency is marked by the yellow lines in the figure.
Since the sunspot model atmosphere has a temperature minimum,
there is a region in between the two yellow lines, where the wave
of frequency 4.5 mHz is evanescent. In the non-magnetic case, the
wave is reflected sharply from this region. In the magnetic case,
the reflection process is more complex. One can observe that the
relative positions of the layer where $c_S = v_A$ and the cut-off
($\omega_c$) layer change with the horizontal distance. Far from
the sunspot axis, where the field is relatively weak, the cut-off
height lies below the $c_S = v_A$ layer. Thus, the waves are
reflected before magnetic effects on the wave path take place.
Note that the trajectories of the waves in the magnetic and
non-magnetic cases are very similar in this region.
As the waves approximate the region close to the sunspot center
(where the magnetic field is large), the position of the $c_S =
v_A$ layer lowers and the fast wave path becomes affected by the
magnetic field. It can be seen from the figure that the red and
white curves start to separate significantly in the region close
to the sunspot axis. The waves in the magnetic case propagate with
larger speeds and overtake the waves in the non-magnetic case.
Note that by introducing the magnetic field, we have changed the
height of the upper turning point of the waves. This may explain
the shift of the positions of the ridges in the magnetic
simulations in Fig.~\ref{fig:kw3} compared to the non-magnetic
simulations in Fig.~\ref{fig:kw}.

Finally, the bottom panel of Fig.~\ref{fig:wkbpath} gives the
propagation paths in the low-frequency case of 3 mHz. In this
case, the position of the cut-off height always lies below the
height of $c_S=v_A$ layer. It means that (at least in our sunspot
model atmosphere) low frequency waves are reflected before their
path is significantly affected by the rapid increase of the
Alfv\'en speed. However, it is important to note that even in this
case, the trajectories of the waves in the magnetic and
non-magnetic models are different. The propagation speed of the
fast magneto-acoustic wave depends on the magnetic field and
always greater than the sound speed.

\section{Quantitative estimates of the travel times and skip distances}

In order to quantify the effects of the magnetic field on the wave
travel times in the eikonal approximation, we calculated the
differences in travel times for the waves in the magnetic and
non-magnetic cases. This calculation is done for waves frequencies
between 3.3 and 4.8 mHz. The resulting travel time differences are
presented in Fig.~\ref{fig:wkbtt} as a function of the wave skip
distance. This calculation is similar to the one reported by
\citet{Moradi+Cally2008}.

The curves shown in Fig.~\ref{fig:wkbtt} were calculated in the
following way. In each case, the waves are launched from their
lower turning point, thus $q=0$, and $p$ is obtained from the
dispersion relation (\ref{eq:disp}). The lower turning point is
located at annuli distance  $x_{\rm A} = 12$ Mm from the sunspot
axis, in order to make the eikonal calculations directly
comparable to the travel time estimates from simulations, as
presented in Fig.~\ref{fig:frequency}.
The vertical location of the lower turning point $z_{\rm A}$
varies in a range from $-$1.5 to $-$10 Mm below the photosphere.
By varying $z_{\rm A}$ and frequency $\omega$, we obtain different
curves in Fig.~\ref{fig:wkbtt}. The deeper is $z_{\rm A}$, the
larger is the value of the skip distance in the horizontal axis of
Fig.~\ref{fig:wkbtt}. Note that to facilitate the comparison
between the numerical simulations and the eikonal approach, we
present the skip distances in Fig.~\ref{fig:wkbtt}  with respect
to the position to the sunspot axis.

The frequency behavior calculated in the eikonal approximation is
in good agreement with the simulations. It proves the reliability
of the eikonal approximation for the full MHD approach.
As in Fig.~\ref{fig:frequency}, the travel time differences are
negative except for positions close to the source location,
meaning that the waves propagate faster in the magnetic case.
There is a strong frequency dependence of the travel time
differences. Fig.~\ref{fig:wkbtt} shows that depending on the
sunspot model and on the skip distance, the maximum values change
from, about, $-$50 sec for the 4.8 mHz wave to $-$10 sec for the
3.3 mHz wave. The agreement between simulations and the eikonal
solution is worse in the $B_{\rm phot}$ = 2.4 kG case, where the
latter gives slightly larger values of the travel time differences
and a different dependence. In all the cases the radius dependence
obtained from the simulations using the Gabor wavelet fit is more
complicated than the skip-distance dependence obtained in the
eikonal approximation.  Note that both simulations and the eikonal
approximation give values of the travel time variations that fit
reasonably well into the range of the observed values in solar
active regions below sunspots \citep[see, \eg\
][]{Kosovichev+etal2000, Zhao+Kosovichev2003,
Couvidat+Birch+Kosovichev2006, Couvidat+Rajaguru2007,
Braun+Birch2008}. However, the frequency dependence in the eikonal
approximation is slightly larger than that obtained from
simulations and observations.

\section{Discussion and Conclusions}

In this paper, we present an investigation of the influence of
magnetic field on helioseismic wave propagation in active regions
below sunspots.
In order to better understand the physics of the waves in complex
magnetic field configurations, we perform a single source
experiment and study the propagation properties of the different
wave modes excited by such a source, as well as the wave travel
times. We have carried out our calculations in a series of sunspot
models with different field strength and for two source positions
located in magnetically and acoustically dominated regions.
The results of our 2D simulations are compared to an asymptotic
solution using the eikonal approximation. The results of both
methods, numerical simulations and eikonal approximation, are
found to be in a good qualitative agreement.
Our approach has allowed us to investigate the effects of the
magnetic field strength in sunspots on the travel-time variations
obtained there from observations using the time-distance
helioseismology techniques.
In addition, the frequency dependence of such measurements has
been studied.

The following conclusions summarize our study:
\begin{itemize}

\item The wave source located immediately below the surface
in a non-magnetic standard solar model excites a mixture of $p$,
$f$ and $g$ mode waves. When the magnetic field is present, the
same source excites a mixture of fast and slow MHD waves. The fast
modes represent an analog of the acoustic $p$ modes, modified by
the magnetic field.

\item In addition to slow MHD waves excited directly by the source,
there is another wave type with low vertical wavelength,
propagating with speed close to that of the fast waves
horizontally across the sunspot. We call them surface
magneto-gravity waves. Several arguments listed in Sect. 4
indicate that these waves are different from either fast or slow
waves. In our simulations these waves propagate below the visible
surface. This makes their detection complicated in spectral
observations.

\item Slow MHD wave modes form an
additional group of ridges in the time-distance diagram with much
lower propagation speed. Their properties depend on the magnetic
field strength of the model, as well as on the height in the
atmosphere where the velocity is measured. The slow mode ridges on
the time-distance diagrams are well isolated and do not affect the
travel time measurements in magnetic regions.

\item  The helioseismic waves below sunspots are speed up by the
magnetic field. The travel times of these waves are shorter by,
about 20--40 seconds compared to the waves in the quiet Sun.
Changing the photospheric magnetic field strength from 0.9 to 2.4
kG produces a variation of the travel times by about 10--15
seconds.

\item Magnetic field produces a strong frequency dependence of the
wave travel times. High frequency waves travel faster than  low
frequency waves. This happens because the propagation path of high
frequency fast waves is modified strongly by the rapid increase of
the Alfv\'en speed in the top layers, since these waves penetrate
higher up in the atmosphere. Low frequency waves are reflected
below the photosphere due to the acoustic cut-off frequency and do
not reach the heights where $v_A > c_S$ and the magnetic fields
produces strong effects. The travel time differences change from
$-20$ to $-35$ seconds between 3 and 5 mHz, in agreement with
observations  \citep[\eg ][]{Rajaguru2008}.

\end{itemize}

The analysis presented here is limited by the 2D approximation. As
a next step in our work we plan to relax this assumption and to
analyze 3D simulations, with a wave excitation by either single or
multiple sources. This will allow us to build a more realistic
picture and to complete our understanding of the interaction of
helioseismic waves with magnetic fields of sunspots.

\acknowledgements  This research has been funded by the Spanish
Ministerio de Educaci{\'o}n y Ciencia through projects
AYA2007-63881 and AYA2007-66502.

%%\aareferences


\begin{thebibliography}{36}
\expandafter\ifx\csname natexlab\endcsname\relax\def\natexlab#1{#1}\fi

\bibitem[{Berenger(1994)}]{Berenger1994}
Berenger, J.~P. 1994, J.\ Comp.\ Phys., 114, 185

\bibitem[{Birch \& Kosovichev(2000)}]{Birch+Kosovichev2000}
Birch, A.~C. \& Kosovichev, A.~G. 2000, Solar Phys., 192, 193

\bibitem[{Braun(1997)}]{Braun1997}
Braun, D.~C. 1997, ApJ, 487, 447

\bibitem[{Braun \& Birch(2008)}]{Braun+Birch2008}
Braun, D.~C. \& Birch, A.~C. 2008, Solar Phys., 251, 267

\bibitem[{Braun \& Lindsey(2000)}]{Braun+Lindsey2000}
Braun, D.~C. \& Lindsey, C. 2000, Solar Phys., 192, 285

\bibitem[{Cally(2005)}]{Cally2005}
Cally, P. 2005, MNRAS, 358, 353

\bibitem[{Cally(2006)}]{Cally2006}
---. 2006, Phil. Trans. R. Soc. A, 364, 333

\bibitem[{Cally \& Goossens(2008)}]{Cally+Goossens2008}
Cally, P.~S. \& Goossens, M. 2008, Solar Phys., 251, 251

\bibitem[{Cameron {et~al.}(2007)Cameron, Gizon, \& Daiffallah}]{Cameron+etal2007}
Cameron, R., Gizon, L., \& Daiffallah, K. 2007, Astron. Nach., 328, 313

\bibitem[{Christensen-Dalsgaard {et~al.}(1996)Christensen-Dalsgaard, Dappen,
  Ajukov, \& \mbox{30 co-authors}}]{Christensen-Dalsgaard+etal1996}
Christensen-Dalsgaard, J., Dappen, W., Ajukov, S.~V., \& \mbox{30 co-authors}.
  1996, Science, 272, 1286

\bibitem[{Couvidat {et~al.}(2006)Couvidat, Birch, \&
  Kosovichev}]{Couvidat+Birch+Kosovichev2006}
Couvidat, S., Birch, A.~C., \& Kosovichev, A.~G. 2006, ApJ, 640, 516

\bibitem[{Couvidat \& Rajaguru(2007)}]{Couvidat+Rajaguru2007}
Couvidat, S. \& Rajaguru, S. 2007, ApJ, 661, 558

\bibitem[{Crouch \& Cally(2003)}]{Crouch+Cally2003}
Crouch, A.~D. \& Cally, P.~S. 2003, Solar Phys., 214, 201

\bibitem[{Duvall {et~al.}(1993)Duvall, Jefferies, Harvey, \&
  Pomerantz}]{Duvall+etal1993}
Duvall, T. L.~J., Jefferies, S.~M., Harvey, J.~W., \& Pomerantz, M.~A. 1993,
  Nature, 362, 430

\bibitem[{Gizon \& Birch(2002)}]{Gizon+Birch2002}
Gizon, L. \& Birch, A.~C. 2002, ApJ, 571, 966

\bibitem[{Gizon {et~al.}(2006)Gizon, Hanasoge, \& Birch}]{Gizon+etal2006}
Gizon, L., Hanasoge, S.~M., \& Birch, A.~C. 2006, ApJ, 643, 549

\bibitem[{Gough(2007)}]{Gough2007}
Gough, D. 2007, Astron.\ Nachr., 328, 273

\bibitem[{Haber {et~al.}(2000)Haber, Hindman, Toomre, Bogart, Thompson, \&
  Hill}]{Habler+etal2000}
Haber, D.~A., Hindman, B.~W., Toomre, J., Bogart, R.~S., Thompson, M.~J., \&
  Hill, F. 2000, Solar Phys., 192, 335

\bibitem[{Hanasoge(2008)}]{Hanasoge2008}
Hanasoge, S.~M. 2008, ApJ, 680, 1457

\bibitem[{Hill(1988)}]{Hill1988}
Hill, F. 1988, ApJ, 333, 996

\bibitem[{Hindman {et~al.}(1997)Hindman, Jain, \& Zweibel}]{Hindman+etal1997}
Hindman, B. W., Jain, R., \& Zweibel, E. G. 1997, ApJ, 476, 392


\bibitem[{Jacoutot {et~al.}(2008)Jacoutot, Kosovichev, Wray, \&
  Mansour}]{Jacoutot+etal2008}
Jacoutot, L., Kosovichev, A.~G., Wray, A., \& Mansour, N.~N. 2008, ApJ,
  684, L51

\bibitem[{Khomenko \& Collados(2006)}]{Khomenko+Collados2006}
Khomenko, E. \& Collados, M. 2006, ApJ, 653, 739

\bibitem[{Khomenko \& Collados(2008)}]{Khomenko+Collados2008}
---. 2008{\natexlab{a}}, ApJ, in press

\bibitem[{Khomenko {et~al.}(2008)}]{Khomenko+etal2008}
Khomenko, E., Collados, M, \& Felipe, T. 2008, Solar Physics, 251, 589

\bibitem[{Kosovichev(1999)}]{Kosovichev1999}
Kosovichev, A.~G. 1999, J. Comput. Appl. Math, 109, 1

\bibitem[{Kosovichev(2002)}]{Kosovichev2002}
---. 2002, Astron.\ Nachr., 323, 186

\bibitem[{Kosovichev \& Duvall(1997)}]{Kosovichev+Duvall1997}
Kosovichev, A.~G. \& Duvall, T. L.~J. 1997, in Solar Convection and
  Oscillations and their Relationship, ed. F.~Pijpers,
  J.~Christensen-Dalsgaard, \& C.~Rosenthal, Vol. 225, Astrophysics and Space
  Science Library (Kluwer Academic Publishers), 241---260

\bibitem[{Kosovichev {et~al.}(2000)Kosovichev, Duvall, \&
  Scherrer}]{Kosovichev+etal2000}
Kosovichev, A.~G., Duvall, T. L.~J., \& Scherrer, P.~H. 2000, Solar Phys., 192,
  159
   
\bibitem[{Lindsey \& Braun(2005a)}]{Lindsey+Braun2005a} 
Lindsey, C. \& Braun, D. C. 2005, ApJ, 620, 1107

\bibitem[{Lindsey \& Braun(2005b)}]{Lindsey+Braun2005b} 
Lindsey, C. \& Braun, D. C. 2005, ApJ, 620, 1118

\bibitem[{McLaughlin {et~al.}(2008)McLaughlin, Ferguson, \&
  Hood}]{McLaughlin+etal2008}
McLaughlin, J.~A., Ferguson, J.~S., \& Hood, A.~W. 2008, Solar Phys., 251, 563

\bibitem[{McLaughlin \& Hood(2006)}]{McLaughlin+Hood2006}
McLaughlin, J.~A. \& Hood, A.~W. 2006, A\&A, 459, 641

\bibitem[{Moradi \& Cally(2008)}]{Moradi+Cally2008}
Moradi, H. \& Cally, P. 2008, Solar Phys., 251, 309

\bibitem[{Parchevsky \& Kosovichev(2007)}]{Parchevsky+Kosovichev2007b}
Parchevsky, K.~V. \& Kosovichev, A.~G. 2007, ApJ, 666, 547

\bibitem[{Parchevsky \& Kosovichev(2008)}]{Parchevsky+Kosovichev2008}
---. 2008, ApJ, submitted

\bibitem[{Rajaguru {et~al.}(2007)Rajaguru, Sankarasubramanian, Wacher, \&
Scherrer}]{Rajaguru+etal2007}
Rajaguru, S. P., Sankarasubramanian, K.,  Wacher, R., \& Scherrer, P. H.
2007, ApJ, 654, L175

\bibitem[{Rajaguru(2008)}]{Rajaguru2008}
Rajaguru, S. P. 2008, ApJ, submitted

\bibitem[{Ruiz Cobo \& del Toro Iniesta(1992)}]{RuizCobo+delToroIniesta1992}
Ruiz Cobo, B. \& del Toro Iniesta, J. C., 1992, ApJ, 398, 375

\bibitem[{Schunker {et~al.}(2005)Schunker, Braun, Cally, \&
  Lindsey}]{Schunker+etal2005}
Schunker, H., Braun, D.~C., Cally, P.~S., \& Lindsey, C. 2005, ApJ, 
   621, L149

\bibitem[{Schunker {et~al.}(2008)Schunker, Braun, Lindsey, \&
  Cally}]{Schunker+etal2008}
Schunker, H., Braun, D.~C., Lindsey, C., \& Cally, P.~S. 2008, Solar Phys., 
   251, 341

\bibitem[{Schunker \& Cally(2006)}]{Schunker+Cally2006}
Schunker, H. \& Cally, P.~S. 2006, MNRAS, 372, 551

\bibitem[{Schwartz \& Stein(1975)}]{Schwartz+Stein1975}
Schwartz, R. A. \& Stein, R. F. 1975, ApJ, 200, 499

\bibitem[{Solanki(2003)}]{Solanki2003}
Solanki, S.~K. 2003, A\&AR, 11, 153

\bibitem[{Zhao \& Kosovichev(2003)}]{Zhao+Kosovichev2003}
Zhao, J. \& Kosovichev, A.~G. 2003, ApJ, 591, 446

\bibitem[{Zhao \& Kosovichev(2006)}]{Zhao+Kosovichev2006}
---. 2006, ApJ, 643, 1317


\end{thebibliography}
\end{document}